\newif
\ifconfver
\confvertrue        
\ifconfver
    \documentclass[10pt,journal,final,twoside]{IEEEtran}
\else
    \documentclass[11pt,draftcls,onecolumn]{IEEEtran}
\fi
\usepackage{xspace,empheq,fancybox,amsmath,amssymb,graphicx,epstopdf,epsfig,subfigure,syntonly,times,amsthm} \usepackage{psfrag,color,bm,array,cite}
\usepackage{url,cite,footnote,xspace,syntonly,algorithm,algorithmic}
\usepackage{verbatim,multirow,slashbox}
\usepackage[T1]{fontenc}

\newtheorem{remark}{\bfseries Remark}
\input{mysymbol.sty}

\newcommand{\diag}{\mathrm{diag}}

\newcommand{\T}{\mathsf{T}}
\begin{document}


\title{Enhancing the Spatio-temporal Observability of Grid-Edge Resources in Distribution Grids}
\author{
\IEEEauthorblockN{Shanny Lin},~\IEEEmembership{Student Member, IEEE}, and \IEEEauthorblockN{Hao Zhu},~\IEEEmembership{Senior Member, IEEE}

\thanks{\protect\rule{0pt}{3mm} {Manuscript received February 14, 2021; revised June 30, 2021; accepted August 7, 2021. This work has been supported by  NSF Grants 1802319, 1807097, and 2130706.}} \thanks{\protect\rule{0pt}{3mm} The authors are with the Department of Electrical \& Computer Engineering, The University of Texas at Austin, Austin, TX, 78712, USA; Emails: {\{shannylin, haozhu\}{@}utexas.edu}.
}}

\markboth{IEEE TRANSACTIONS ON SMART GRID (ACCEPTED)}
{Lin \MakeLowercase{\textit{et al.}}: Enhancing the Spatio-temporal Observability of Grid-Edge Resources in Distribution Grids}
\renewcommand{\thepage}{}
\maketitle
\pagenumbering{arabic}

%
\begin{abstract} 
Enhancing the spatio-temporal observability of distributed energy resources (DERs) is crucial for achieving secure and efficient
operations in distribution grids. This paper puts forth a joint recovery framework for residential loads by leveraging the complimentary strengths of heterogeneous measurements in real time. The proposed framework integrates low-resolution smart meter data collected at every load node with fast-sampled feeder-level measurements from limited number of distribution phasor measurement units. To address the lack of data, we exploit two key characteristics for the loads and DERs, namely the sparse changes due to infrequent activities of appliances and electric vehicles (EVs) and the locational dependence of solar photovoltaic (PV) generation.  Accordingly, meaningful regularization terms are introduced to cast a convex load recovery problem, which will be further simplified to reduce the computational complexity. The load recovery solutions can be utilized to identify the EV charging events at each load node and to infer the total behind-the-meter PV output. Numerical tests using real-world data have demonstrated the effectiveness of the proposed approaches in enhancing the visibility of  grid-edge DERs. \end{abstract}



\section{Introduction}\label{sec:intro}
Distribution grids have witnessed an increasing penetration of distributed energy resources (DERs) at the grid-edge. Electrical vehicles (EVs) and renewable generation owned by residential customers and aggregators lead to growing dynamics in distribution systems. High-resolution dynamic profiles of residential loads have been advocated for validating EV charging commands \cite{Niddodi2019ev} and photovoltaic (PV) inverter control settings \cite{Jacobs2019pv}. In addition, the instantaneous total heating/cooling loads and solar PV generation within a feeder are useful for designing protection systems and demand response programs \cite{kabir2020joint,ledva2020separating}.
Therefore, it is important to enhance the spatio-temporal observability of grid-edge DERs for efficient and secure grid operations. 

Nonetheless, distribution systems traditionally lack in sensing infrastructure and real-time observability. To address this issue, different types of distribution sensors have been advocated recently. At the feeder level, line flow meters and distribution phasor measurement units (D-PMUs) \cite{Meier2014} can provide fast power measurements of high quality, but only at selected locations. Meanwhile, almost all residential homes are equipped with smart meters that can collect the electricity consumption data over intervals of 15 minutes to an hour \cite{AMI}. Hence, high resolution D-PMU data can suffer from low spatial diversity, while the ubiquitous smart meter data lack in the temporal resolution. These two types of measurements provide complimentary strengths, and the question opens up on how to jointly utilize them both to enhance the visibility of  grid-edge DERs.

The present work is highly related to the distribution state estimation (DSE) problem. 
Traditional DSE methods \cite{lu1995dsse,baran1995branch} have focused on  the static setting and identified the lack of measurements as the main challenge. To tackle this, more recent DSE methods have considered the coupling between two consecutive time instances \cite{bhela2018enhancing} and the spatial correlation among voltage and power data \cite{donti2020matrix}. All of these approaches are not directly applicable to the case of different measurement time resolutions. Motivated by the availability of fast D-PMU data, dynamic DSE methods have been developed by building on Kalman filtering {\cite{carquex2018state,zhao2019robust}} or prediction-correction method \cite{song2020dynamic}, {including a recent extension to asynchronous data in \cite{cavraro2019dynamic}}. These recursive update based methods can well track slowly varying system changes but may fail to cope with fast start/end events that are common for residential appliances and DERs. {In addition, pseudo measurements using load forecast have been popularly considered for DSE, such as the Bayesian estimation approach in \cite{schenato2014bayesian}. Nonetheless, it is very difficult to predict appliance activities from historic data.} {See \cite{Primadianto2017review,wang2019distribution}} 
for comprehensive reviews of the DSE literature. 
{Another line of related work is the load disaggregation for inferring behind-the-meter (BTM) PV output. However, these approaches typically require prior knowledge such as historical load data for a subset of customers or the statistical models for the total consumption; see e.g.,  \cite{kabir2020joint,cheung2018behind} and references therein. Such information can be difficult to obtain while most of these approaches only utilize slow smart meter data. 
Thus, the problem still remains on how to effectively incorporate dynamic data streams of various resolutions to enhance the observability of grid-edge DERs.} 

The goal of this paper is to develop a joint spatio-temporal DER inference framework by using the respective strengths of  D-PMU and smart meter data. Both types of data are first modeled as linear transformation of the unknown power demands. This model builds upon linearizing the multiphase power flow equations and generalizes our earlier work {\cite{lin2019enhancing}} for lossless systems. To tackle the lack of measurements, two key underlying characteristics of the spatio-temporal load demand 
are exploited. First, the on-off activities of household appliances and EVs typically occur \emph{infrequently}, leading to \emph{jointly sparse changes} of both active and reactive powers. Second, renewable generation such as rooftop solar PV demonstrates a strong \textit{spatial correlation} within a feeder, giving rise to a \emph{low-rank} component in the active power demand. 
{Note that the correlation of localized PV outputs here is different from earlier work \cite{batra2018transferring,miyasawa2019energy} on low-rankness of the load data. These approaches exploit the relation among various household appliances for disaggregating the total house load consumption.} To promote the low-rank plus sparse change feature, group L1-norm \cite{yuan2006glasso} 
and nuclear norm \cite{candes2011robust,wright2009robust,recht2010guaranteed}  are introduced as regularization, leading to a convex formulation. 
{The low-rank component is further simplified into a rank-one PV component to reduce the computational complexity associated with nuclear norm. Numerical tests using real-world load and DER data have been conducted to demonstrate the effectiveness of the joint inference formulations in identifying EV events and estimating the total 
{BTM} solar output in a feeder. We have observed that a few D-PMUs can significantly improve the DER visibility than only using smart meter data, while the presence of periodic appliance activities can affect the performance of DER inference.}   

The main contribution of this work is summarized here. First, we consider the inference problem for monitoring the dynamics of residential DERs that can incorporate distribution system measurements of different time resolutions and spatial availability in Section \ref{sec:sys_mod}. Second, the proposed spatio-temporal inference framework has exploited the underlying low-rank and sparse-change characteristics to improve the recovery performance of residential DERs in Section \ref{sec:net_mod}. Last, the proposed methods have been successfully applied to crucial grid-edge monitoring tasks such as identifying the exact timing of EV events and estimating the 
BTM solar generation in Section \ref{sec:sim}. Conclusions and future directions are drawn in Section \ref{sec:con}.

\textit{Notation:} Upper (lower) boldface symbols stand for matrices (vectors); $(\cdot)^{\T}$ stands for matrix transposition; $(\cdot)^*$ complex conjugate; $\|\cdot\|_*$ denotes the matrix nuclear norm; $\|\cdot\|_g$ the group L1-norm; $\|\cdot\|_2$ the L2-norm; $\otimes$ denotes the Kronecker product; $\mathbf I$ ($\mathbf 1$) stands for the identity matrix (all-one vector) of appropriate size; 
$\diag(\cdot)$ represents the diagonal matrix; {and symbols using $\check{\cdot}$ indicate vectors of all non-reference bus quantities}.

\section{System Model}
\label{sec:sys_mod}

\begin{figure}[tb] 
\centering
\vspace{-3pt}
\includegraphics[width=.8\linewidth,clip = true]{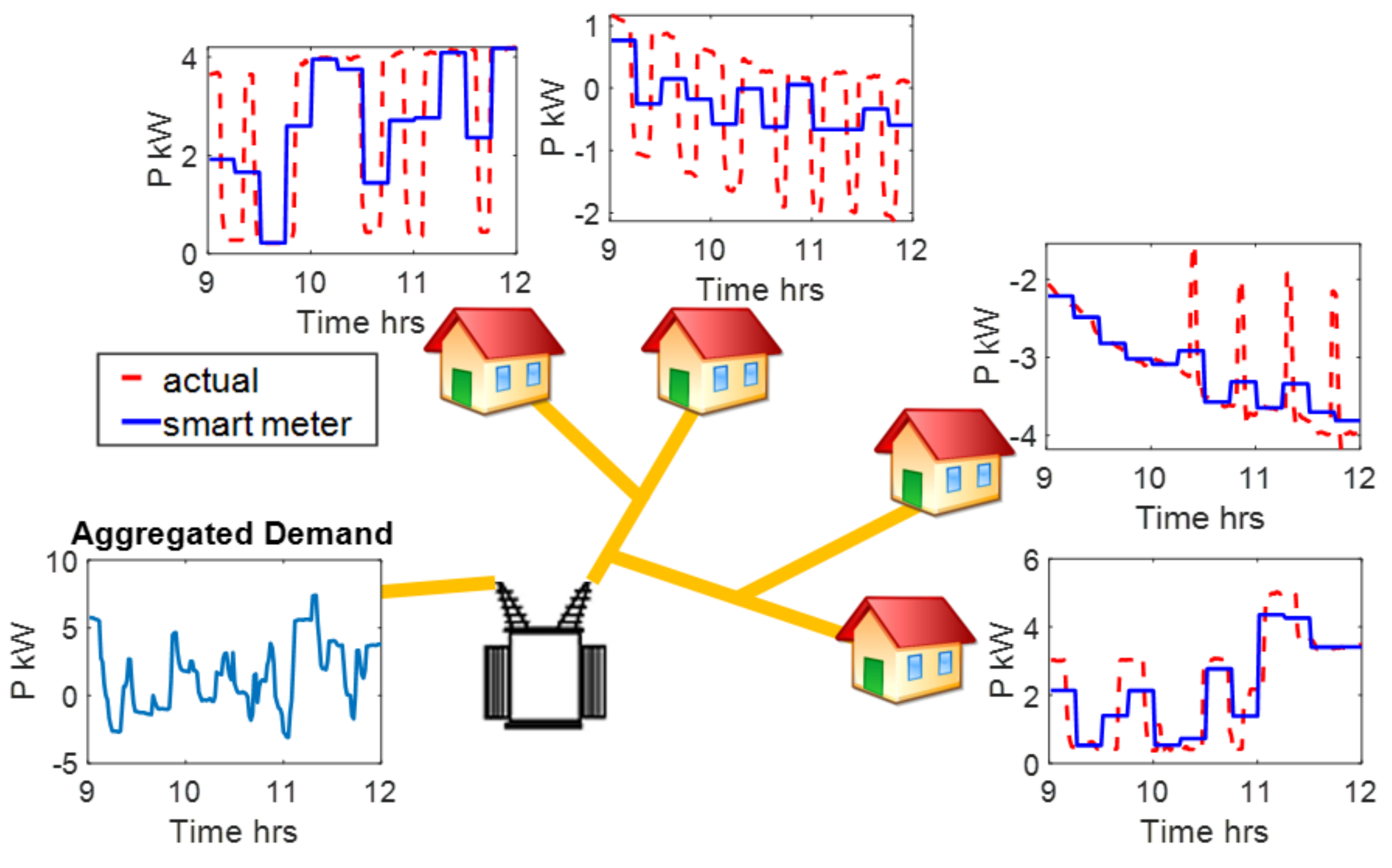}
\vspace{-4mm}
\caption{Overview of a simple distribution feeder system with four residential loads and various types of power measurements.} \label{fig:sys_overview}
\end{figure}

Consider a multiphase distribution feeder with residential loads at the feeder ends shown by Fig.~\ref{fig:sys_overview}. We are interested in  the spatio-temporal complex load matrix $(\bbP+\mathsf{j}\bbQ) \in \mathbb{C}^{N\times T}$, or the real-valued version {$\bbX \coloneqq [\bbP; \bbQ] \in \mathbb R^{2N\times T}$}, where $N$ denotes the number of single-phase loads and $T$ the number of time slots. The temporal resolution represents the fastest sampling rate of all available measurements. For example, the sampling rate can be at the sub-second scale which corresponds to the time resolution of D-PMUs \cite{Meier2014}. Without loss of generality (Wlog), this paper uses the fast sampling rate at every minute. 

Two types of measurement data will be mainly utilized, namely the smart meter data for each load and the aggregated feeder-level measurements. Residential houses are equipped with smart meters to record the average electricity consumed over intervals of 15 minutes to an hour \cite{AMI}. Here we assume both real and reactive powers are collected every 15 minutes. Using this down-sampling rate to obtain  $T_s = T/15$ samples, one can form the smart meter data matrix of size $2N\times T_s$ as   
\begin{align} \label{sm_eqn}
\bbGamma = \begin{bmatrix}
\bbGamma^P \\
\bbGamma^Q
\end{bmatrix} = \begin{bmatrix}
\bbP \\
\bbQ
\end{bmatrix} \bbA + \bbN^\Gamma
\end{align} 
where the $T\times T_s$ matrix $\bbA = \frac{1}{15}\bbI \otimes \mathbf{1}$ is the time averaging operator, while $\bbN^\Gamma$ denotes the measurement noise matrix. {We extend \eqref{sm_eqn} to the case of asynchronized sampling, by using a node-specific averaging $\bbA_n$ according to its sampling intervals. As shown by the numerical tests later on, this generalization can be easily included by the rest of analysis as the measurement model is still linear.}

At the feeder level, D-PMUs or other meters installed on power lines can measure quantities such as  aggregated demand  at a fast rate. To model the feeder-level data, we briefly introduce the multiphase ac power flow equations; see e.g., \cite {kersting2007} for a comprehensive overview. Let the substation or feeder head be the reference bus 0, and use $N_b$ denote the number of other buses (PQ buses). Wlog, all the buses are three-phase connected, and for simplicity, only wye-connected single-phase loads are considered. Let vectors {$\check{\bbs} \in \mathbb{C}^{3N_b}$ and $\check{\bbv} \in \mathbb{C}^{3N_b}$} collect the complex power injections and phase-to-ground voltages at every phase of the PQ buses. The power injection {$\check{\bbs}$} is quadratically related to the voltage {$\check{\bbv}$}, as   {
\begin{align}\label{AC_eqn}
\check{\bbs} &= \textsf{diag}(\check{\bbv})\check{\bbi}^* = \textsf{diag}(\check{\bbv})\left(\bbY_{L} [\bbv_0; \check{\bbv}]\right)^* 
\end{align}
}where {$\bbv_0 \in \mathbb C^3$ is the given reference voltage vector at feeder head, while matrix $\bbY_L \in \mathbb C^{3N_b\times 3(N_b+1)}$ is the submatrix of the full admittance (Ybus) matrix $\bbY = [\bbY_0; \bbY_L] \in \mathbb C^{3(N_b+1)\times 3(N_b+1)}$ that consists of rows of all non-reference PQ buses.} 
Similarly, the power injection to the reference bus {$\bbs_0\in \mathbb C^3$}, or the total aggregated power demand, becomes {
\begin{align}
\bbs_0 = \textsf{diag}(\bbv_0)\bbY_{0}^* [\bbv_0; \check{\bbv}]^* 
\label{s0_eqn}
\end{align} 
where $\bbY_0 \in \mathbb C^{3\times 3(N_b+1)}$} is the submatrix of Ybus with rows corresponding to the reference bus. 
In general, $\bbs_0$ and the power/current flow elsewhere in the feeder are fully determined by the voltage profile {$\check{\bbv}$}. 

To simplify the solution of {$\check{\bbv}$}, one can adopt linearized power flow approximations such as the fixed-point method  (FPM) developed in \cite{bernstein2018load,gatsis2018convergence}. 
Consider the linearized version of  \eqref{AC_eqn} given by  {
\begin{align}
\tilde{\bbv} = \bbM\check{\bbx} + \check{\bbw}
\label{vlin_eqn}
\end{align}
where $\check{\bbx} \coloneqq [\textsf{Re}\{\bbs\}; \textsf{Im}\{\bbs\}] \in \mathbb R^{6N_b}$ 
has the full nodal injection and $\check{\bbw} \in \mathbb C^{3N_b}$} is the complex voltage solution under zero-loading condition. Basically, the FPM-based approximation is  the linear interpolation of {$\check{\bbw}$} and another voltage solution (hence the term fixed-point). Substituting the linearized voltage model \eqref{vlin_eqn} into \eqref{s0_eqn} yields {
\begin{align}
\tilde{\bbs}_0 = \textsf{diag}(\bbv_0)\bbY_{0L}^*\bbM^*\check{\bbx}
 \label{s0_lin_eqn}
\end{align}
where $\bbY_{0L} \in \mathbb C^{3\times 3N_b}$ is the submatrix of $\bbY_0 = [\bbY_{00},\bbY_{0L}]$ formed by the PQ-bus columns only.} Note that the additional offset term  is canceled in \eqref{s0_lin_eqn} so that $\tilde{\bbs}_0=\mathbf{0}$ under the zero-loading condition. Wlog, let {$\check{\bbx}$} 
 include only the load nodes of non-zero injection, and thus it 
{reduces to a ${(2N\times 1)}$ vector $\bbx$} which matches the dimension of matrix $\bbX$. By stacking \eqref{s0_lin_eqn} across $T$ time slots, we obtain the measurement model for the aggregated demand as {
\begin{align}
\bbZ = \begin{bmatrix}
	\bbZ^P \\
	\bbZ^Q
\end{bmatrix} 
= \bbH \bbX + \bbN^Z \in \mathbb R^{6\times T}
\label{agg_eqn}
\end{align}
}where $\bbN^Z$ is the measurement error matrix. 
\begin{remark}[Measurement types and accuracy] \label{rmk:agg}
The FPM-based measurement model \eqref{agg_eqn} can encompass a variety of feeder-level quantities in addition to aggregated power, such as the data on line power/current flow and nodal voltage phasor/magnitude as measured by D-PMUs; see more discussions in \cite{Primadianto2017review}. 
{Intuitively, the larger the amount of available data, the better the performance a recovery algorithm would attain. For example, our numerical experiences (not included due to space limit) confirm that the proposed joint recovery framework using both D-PMU and smart meter data clearly outperforms the recovery using the latter only.} As for the accuracy of the FPM-based linear approximation, it depends on the fixed power flow solution used to form $\bbM$, as shown in \cite{bernstein2018load,gatsis2018convergence}. Ideally, this fixed point should closely approach the actual power flow solution. Using the latest smart meter data, one can 
{update matrix $\bbM$ of \eqref{s0_lin_eqn}} for
{improved accuracy of linearization. As the aggregated demand model \eqref{agg_eqn} would still be linear in $\bbX$, this time-varying $\bbM$ scenario can be easily considered by the rest of analysis as shown by our numerical tests later on.} 
\end{remark}

Even though encompassing various measurements, the problem of recovering $\bbX$ is expected to be severely under-determined.
For the measurement models in \eqref{sm_eqn} and \eqref{agg_eqn}, the total number of equations equals $(2N T_s+6T)$ which is much smaller than $(2NT)$, the number of unknowns. This motivates one to exploit the special characteristics of the load matrices for improving the recovery performance. 

\section{Monitoring DERs via Load Recovery} \label{sec:net_mod}

\begin{figure}[t!]
\centering
\includegraphics[width=.8\linewidth,trim=0 0 20 20, clip]{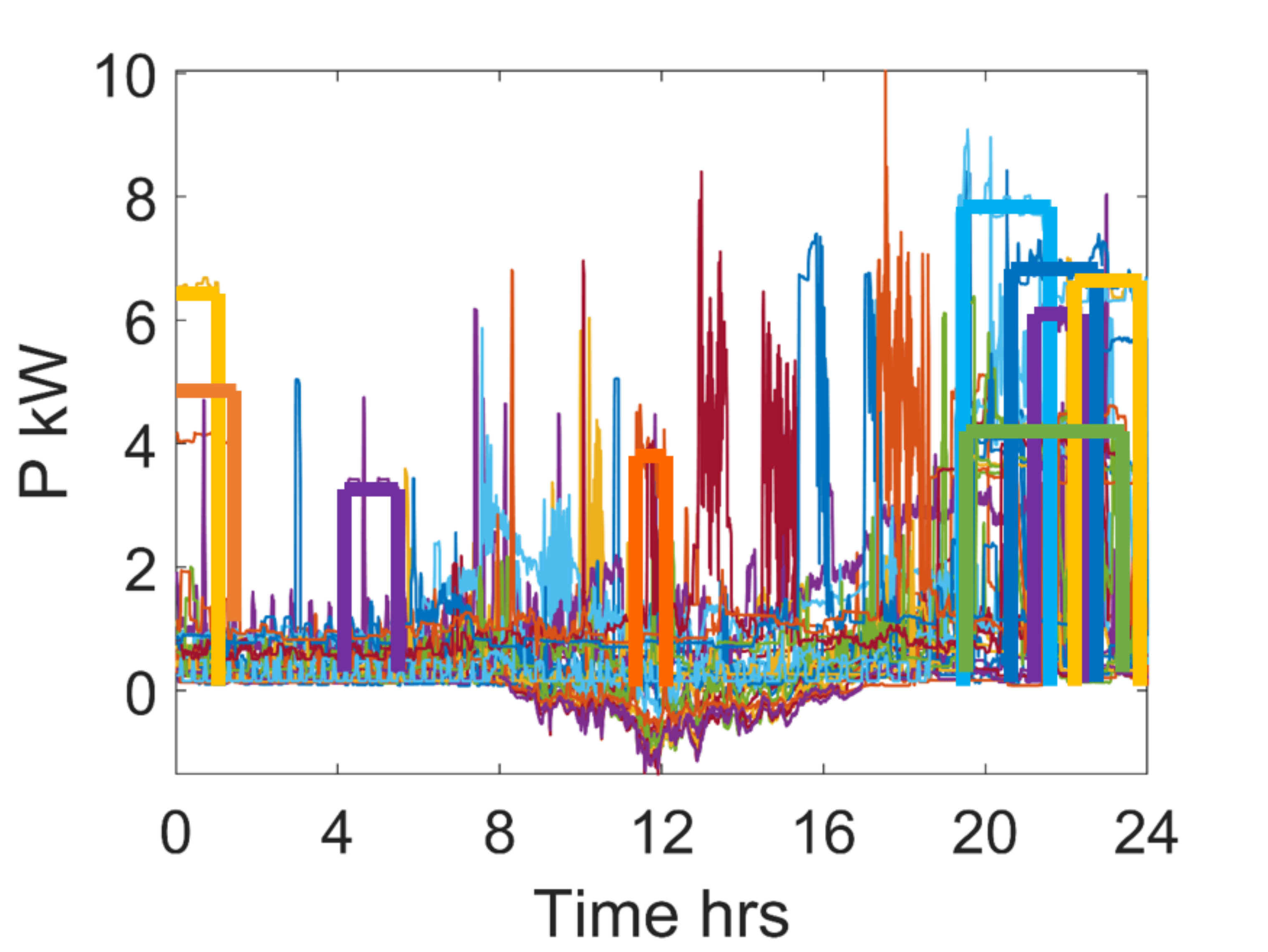}
\caption{Residential load profiles of active power from Pecan Street's Dataport \cite{pecan} for 30 houses, 15 of which have rooftop PVs. }
\label{fig:P_observed}
\end{figure}

We first discuss the characteristics of residential load matrices that are amendable for formulating the load recovery problem. Typically, residential load profiles consist of the base loading and household appliance activities; see e.g., \cite{proedrou2021comprehensive}. Hence, each residential load can be approximated by the aggregation of its own rectangular waveforms that are independent from other locations. DERs such as EVs are operated as household appliances of large power ratings, and thus their load profiles also follow the rectangular waveform condition.   For example, Fig.~\ref{fig:P_observed} plots the active power profiles of 30 actual residential homes, of which the large night-time rectangular waveforms correspond to EV charging activities. A majority of appliance activities are observed to occur \textit{infrequently} throughout the day, and have a non-unity inductive power factor \cite{kuzlu2014load}. Hence, they can be represented as \emph{sparse changes} in the household load profile, in both active and reactive power.  Notice that this sparse-change characteristic was also exploited in \cite{bhela2018enhancing} for static distribution system state estimation. {Some periodic appliances such as HVACs may violate this condition, as discussed later in Remark \ref{rmk:hvac}.} 

To this end, let $\bbS^P, \bbS^Q \in  \mathbb{R}^{N\times T}$ denote the respective \textit{sparse-change} components of $\bbP$ and $\bbQ$ corresponding to infrequent appliance activities. These two matrices will have piece-wise constant rows and the respective consecutive time differences of the entries, denoted by matrices $\bbD^P$ and $\bbD^Q$, are sparse. In other words, $\bbS^P$ becomes sparse under the linear transformation $\bbS^P = \bbD^P\bbU$  where $\bbU \in \mathbb{R}^{T\times T}$ is the upper triangular matrix of all ones; and similarly for $\bbS^Q = \bbD^Q\bbU$. This way, the number of non-zero entries $D_{n,t}^P$ and $D_{n,t}^Q$, that indicate the start/end of appliance activities, is much less compared to the total number of entries. Furthermore, each pair $\{D_{n,t}^P,D_{n,t}^Q\}$ is \textit{jointly sparse},  as any appliance activities could manifest in \textit{synchronized} changes of active/reactive power demand. This joint sparsity will be explored later for the load recovery problem.

Different from residential appliances,  DERs like rooftop PVs would demonstrate correlated behaviors among nearby locations. Typically, PV generation output depends on localized factors such as solar irradiance and weather conditions \cite{sharma2011predicting}. Hence, the active power generated by rooftop PVs within a neighborhood would share the same temporal pattern. This can be observed in Fig.~\ref{fig:P_observed} where half of the houses have PV installation and share a negative day-time component. We denote this underlying spatial correlation in $\bbP$ by a low-rank component $\bbL \in \mathbb{R}^{N\times T}$ of highly correlated rows. It will be further considered as a rank-one component later on. 
{It is worth mentioning that the proposed low-rank plus sparse change model includes typical residential appliances and thus it encompasses general types of electricity consumption or generation in residential loads. For the interest of grid-edge monitoring, it will be mainly used for demonstrating the applications to EV event identification and BTM solar PV recovery later on.}

\begin{remark}[Reactive power support]\label{rmk:var}
	Reactive power is assumed to have no spatial correlation which follows from the unity power factor setting under the IEEE Standard 1547-2018 \cite{8332112}. This is currently the most common DER operation mode which does not provide grid reactive power support. Nonetheless, certain reactive power functions from DERs can be included, such as constant (non-unity) power factor or general Watt-Var curve where $\bbQ$ would share the same temporal pattern as $\bbP$. Hence, for general DER modes, we can slightly modify the data model to have the low-rank component also present in $\bbQ$.
\end{remark}

Under the two load characteristics, the active power matrix can be decomposed into $\bbP = \bbL + \bbD^P \bbU$.  This representation mimics the one in robust principal component analysis (RPCA), namely a low-rank plus sparse model {\cite{candes2011robust}}. To better reveal this model, we use the linear transformation that essentially computes the differences between consecutive columns (time instances) and obtain $\bbP = (\bbK+\bbD^P)\bbU$ where $\bbK := \bbL\bbU^{-1}$ is also of low rank. This is because of the temporal pattern in $\bbL$ leads to strong  correlation of the differences as well. Similarly, we have $\bbQ = \bbD^Q\bbU$ which could include a low-rank component (cf. Remark \ref{rmk:var}). Clearly, the measurements in \eqref{sm_eqn} and \eqref{agg_eqn} are still linearly related to the unknown matrix components, namely $\bbK$, $\bbD^P$ and $\bbD^Q$.  

The problem now boils down to recovering the low-rank and sparse matrix components, for which useful regularization terms can be introduced. The nuclear norm regularization is widely adopted in low-rank matrix recovery such as RPCA \cite{candes2011robust}, subspace learning \cite{wright2009robust}, and collaborative filtering \cite{recht2010guaranteed}. It is defined as the sum of  matrix's singular values, given by
\begin{align}
\textstyle \|\bbK\|_{*} \coloneqq \sum^N
_{i=1} \sigma_i(\bbK)
\label{eqn:nuc_norm}
\end{align}
where $\sigma_i(\cdot)$ denotes the \textit{i}-th largest singular value. The nuclear norm is a convex norm since it is the dual function of the matrix spectral norm; see \cite[pg. 637]{Boyd}. As for the sparse components, the popular L1-norm regularization defined as the sum of entry-wise absolute values is useful. The L1-norm is a tight convex relaxation of the L0 pseudo-norm (the total number of nonzero entries), and it has been shown to efficiently attain the sparse signal representation as widely used in compressed sensing and sparse signal recovery \cite{ct06tit,Donoho2006}. Entries of $\bbD := [\bbD^P;\bbD^Q]$ are not only sparse but also \textit{jointly sparse}, because each appliance activity induces simultaneous changes in $\{D_{n,t}^P,D_{n,t}^Q\}$. Accordingly, we use the following \emph{group} L1-norm definition: 
\begin{align}
\|\bbD\|_{g} \coloneqq	\sum_{n=1}^N \sum_{t=1}^T \left\|[D^P_{n,t}; D^Q_{n,t}]\right\|_2.
	\label{eqn:gLasso}
\end{align}
This group L1-norm can be thought of as the L1-norm for the jointly sparse pairs, and has been popularly used to promote sparsity at the group level   \cite{yuan2006glasso}. Note that it is also convex as {each summand term is the convex L2-norm}.

Based on the two norm terms, we can formulate the following recovery problem:
\begin{subequations}\label{eqn:multi_batch}
\begin{align}
& \min_{\bbK,\bbD} && ~~~~\|\bbK\|_{*} + \lambda \|\bbD\|_g \\
& \textrm{s. to} && - \boldsymbol{\mathcal{E}} \leq \bbGamma - \begin{bmatrix}
\bbK + \bbD^P \\
\bbD^Q
\end{bmatrix} \bbU \bbA \leq \boldsymbol{\mathcal{E}} \label{eq:mbb}\\
& ~ && - \bbE \leq \bbZ - \bbH
\begin{bmatrix}
\bbK + \bbD^P \\
\bbD^Q
\end{bmatrix} \bbU \leq \bbE \label{eq:mbc}
\end{align}
\end{subequations}
where the coefficient $\lambda > 0$ can balance between the two norms, while $\boldsymbol{\mathcal{E}}$ and $\bbE$ contain the pre-determined entry-wise error bounds for the measurements. Measurement error can be bounded based on the metering accuracy or the model mismatch. If only measurement noise is considered, the bounds are set according to the largest metering error level for each datum. As Remark \ref{rmk:agg} points out, the linearized model \eqref{agg_eqn} for $\bbZ$ would suffer from approximation error, which is included by the maximum deviation bounds in $\bbE$. Other types of error metrics may be used as well, such as {the} Frobenius norm which is preferred for Gaussian distributed error terms. For simplicity, the entry-wise maximum error bounds \eqref{eq:mbb}-\eqref{eq:mbc} have been used as linear constraints. Additional physics-based constraints such as non-negative $\bbQ$ may be considered as well.

\begin{remark}[Choice of $\lambda$]\label{rmk:lambda}
{	
To select  $\lambda$, earlier work \cite{candes2011robust,wright2009robust} suggests to scale it with the matrix dimension as $\mathcal{O}(1/ \sqrt{T})$ for $T \gg N$.  We have followed this scaling with additional parameter tune-up.  As mentioned earlier, this hyper-parameter balances between the two matrix norms. 
Therefore, the value of $\lambda$ would control the sparsity level of the recovered $\bbD$, or equivalently, the frequency of load changes. For very large $\lambda$, the recovered load profiles will be entirely contained in $\bbL$ with all-zero $\bbD$, and thus all loads are completely correlated with minimal individual activities. Similarly, under very small $\lambda$, there will be low correlation among the recovered profiles. Without the actual profiles available for validation, we have chosen the value of $\lambda$ according to the typical occurrence of load change activities after adopting the $\mathcal{O}(1/ \sqrt{T})$ scaling. Our numerical experiences suggest that the recovery performance is pretty robust to the $\lambda$ choice. Thus, the parameter tuning step by matching the frequency of load changes can attain a satisfactory performance for the load recovery problem. }
 \end{remark}

Since both norms are convex functions, \eqref{eqn:multi_batch} is a convex problem. The computational time associated with the nuclear norm, however, grows very quickly with the problem dimension as the resultant semidefinite program having a cubic computational complexity and a quadratic memory requirement \cite{toh2010accelerated}. To tackle this issue, we can simplify problem \eqref{eqn:multi_batch} by assuming that the (relative) solar capacity is known for each house. This information can be obtained from customer reporting or historical data analysis by solving \eqref{eqn:multi_batch}. Hence, the solar PV output is represented using a rank-one matrix as $\bbK = \bbu \bbv^\T$, where each entry of $\bbu$ is proportional to the solar capacity per house. This way, the problem now becomes a linear estimation one for $\bbv$ and $\bbD$, as given by
\begin{subequations}\label{eqn:multi_batch2}
	\begin{align}
\min_{\bbv,\bbD} &~~~~ \frac{1}{2}\|\bbv\|_2^2 + \lambda \|\bbD\|_g \\
\textrm{s. to} & - \boldsymbol{\mathcal{E}} \leq \bbGamma - \begin{bmatrix}
			\bbu \bbv^\T + \bbD^P \\
			\bbD^Q
		\end{bmatrix} \bbU \bbA \leq \boldsymbol{\mathcal{E}} \\
		& - \bbE \leq \bbZ - \bbH
		\begin{bmatrix}
			\bbu \bbv^\T + \bbD^P \\
			\bbD^Q
		\end{bmatrix} \!\bbU \leq \bbE 
	\end{align}
\end{subequations}
{where the squared L2-norm of vector $\bbv$ is used to replace the nuclear norm regularization. This change follows from 
	an alternative characterization of the nuclear norm of a rank-one matrix \cite{recht2010guaranteed}. We will implement this simplified formulation for a large number of nodes.  
	
\begin{remark} [BTM solar disaggregation] \label{rmk:BTM} The solar PV  temporal pattern can be estimated by $\hhatbbrho^T = \hhatbbv^\T \bbU$, where $\hhatbbv$ represents either the solution to \eqref{eqn:multi_batch2} or the first right singular vector of the estimated $\bbK$ by \eqref{eqn:multi_batch}. The estimated $\hhatbbrho$ can be used to recover the total BTM solar output from the aggregated measurement $\bbZ^P$. To this end, the aggregated measurement per phase $\phi$ can be modeled as {$\bbz^P_\phi \cong (\alpha \mathbf{1} + \beta \hhatbbrho) + \bbU^\T \bbd_\phi^P$}, where the unknown coefficients $\alpha$ and $\beta$ transforms the solar pattern to the actual PV output while vector $\bbd_\phi^P$ is the total sparse load change for phase $\phi$. Given $\hhatbbrho$ and solar outputs being zero at non-daytime hours, one can use this linear model of $\bbz^P_\phi$ to estimate the three unknowns. The estimated $(\alpha \mathbf{1} + \beta \hhatbbrho)$ can separate the total solar outputs on phase $\phi$.  \end{remark}

\begin{remark}[Periodic appliances]\label{rmk:hvac} Appliance activities are assumed uncorrelated among houses, which may not hold for certain periodic loads such as HVACs. Residential HVAC consumption highly depends on local weather conditions, and thus its periodicity is really similar for co-located houses. As a result, the HVAC demand component can be thought of as the Fourier series expansions of a same periodicity, which is of low-rank as well. Theoretical results for the RPCA framework \cite{candes2011robust} imply that accurate recovery of the unknown matrix requires that the low-rank and sparse components do not overlap. Loosely speaking, the low-rank $\bbK$ is not sparse while the sparse $\bbD$ is not of low rank. Hence, the periodic appliance component, both sparse and low-rank, would be present in the recovered temporal pattern $\hhatbbrho$, as demonstrated by numerical tests soon. A band-pass filter could resolve this issue by filtering out the signal of the HVAC periodicity (if known), which is out of scope for the present work.   
\end{remark}

\section{Numerical Tests} \label{sec:sim}

We have conducted numerical studies using real-world load datasets to evaluate the effectiveness of both recovery approaches  \eqref{eqn:multi_batch} and \eqref{eqn:multi_batch2}. We have constructed two multi-phase test systems by modifying the {R2-12.47-3 feeder} from the GridLab-D taxonomy feeder deposit \cite{taxfeeder}. 
We have one small {33-node} feeder that can host 30 residential houses and another {111-node} one with 100 residential houses. In both systems, half of the houses are installed with solar PVs, and the minute-level active power matrix $\bbP$ has been obtained from the PecanStreet's Dataport \cite{pecan} for homes located in Austin, TX. {Active power data obtained for the months of December and July have been used to generate the winter and summer datasets, respectively}. An example daily load profile for the 30 houses during a winter day is shown in Fig.~\ref{fig:P_observed}. The reactive power matrix $\bbQ$ was formed by randomly selecting a power factor in the range $[0.9,~0.95]$. In addition, both $\bbP$ and $\bbQ$ were scaled to match the original loading of the R2-12.47-3 feeder. 

To generate the D-PMU measurement data, we have run the nonlinear multiphase power flow simulation using GridLab-D \cite{gridlabd} for both systems. Random noise was added to the simulated data to form the measurement $\bbZ$. 
We used the uniform distribution in the range of $[-0.02\%,0.02\%]$ error for generating the noise in each D-PMU power measurement, per the total vector error of $\pm0.01\%$ in a typical D-PMU datasheet \cite{PSL}. 
{The first D-PMU is always placed at the feeder head. Additional D-PMUs are placed at lateral heads to aggregate different subsets of down-stream loads. Intuitively, more D-PMUs can provide new information while the location of D-PMUs matters too. For example, D-PMUs installed at load nodes at the end of feeder are not as useful as those installed at the head of feeder or laterals. Hence, we have picked the D-PMU locations to aggregate as many load nodes as possible.} As for the smart meter data, it was obtained by the averaging over every 15-minute window as in \eqref{sm_eqn}. We used the $\pm0.2\%$ accuracy from the ANSI C12.20 Standard \cite{ANSI} to randomly generate uniformly-distributed additive noise for every smart meter measurement.

\begin{figure}[t!]
	{
		\centering
		\includegraphics[width=.8\linewidth]{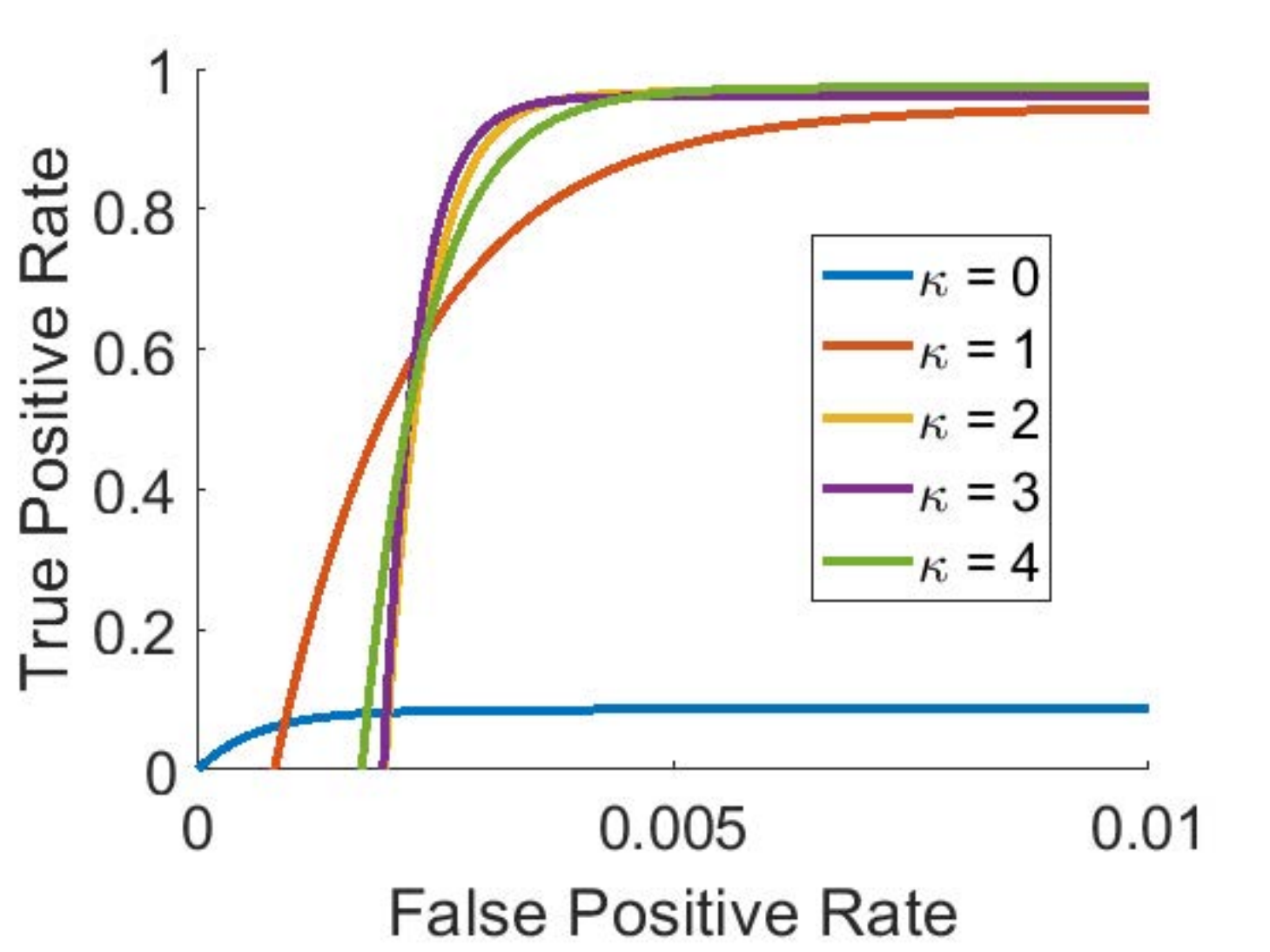}
		\centerline{(a)}
		\includegraphics[width=.8\linewidth]{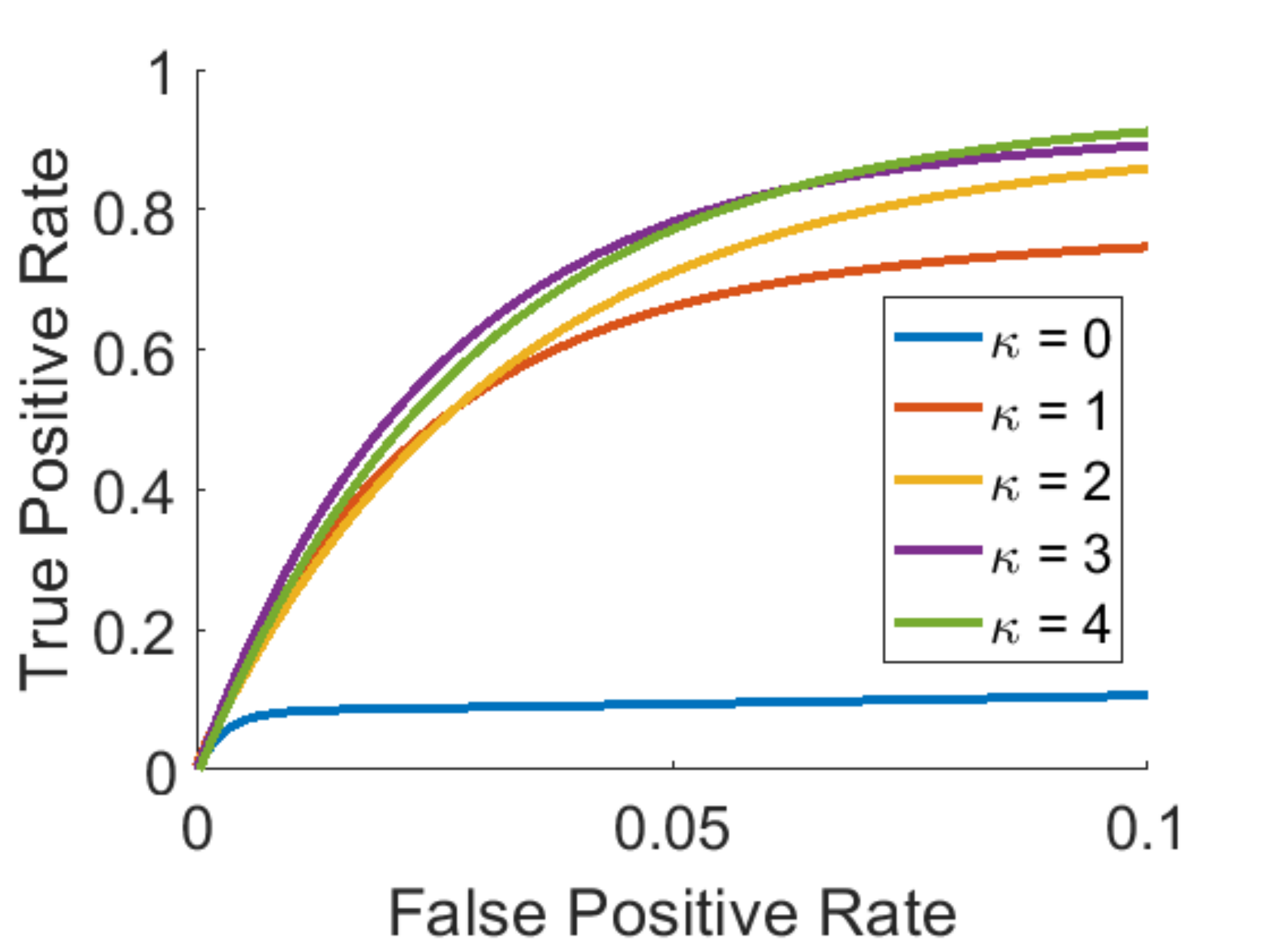}
		\centerline{(b)}
		\caption{{Receiving operating curves (ROCs) of EV detection in Test Case 1 for the (a) winter and (b) summer night-time datasets under different number of D-PMUs.}}
		\label{fig:roc_30house}}
\end{figure}

\begin{figure}[t!]
	{
		\centering
		\includegraphics[width=.8\linewidth]{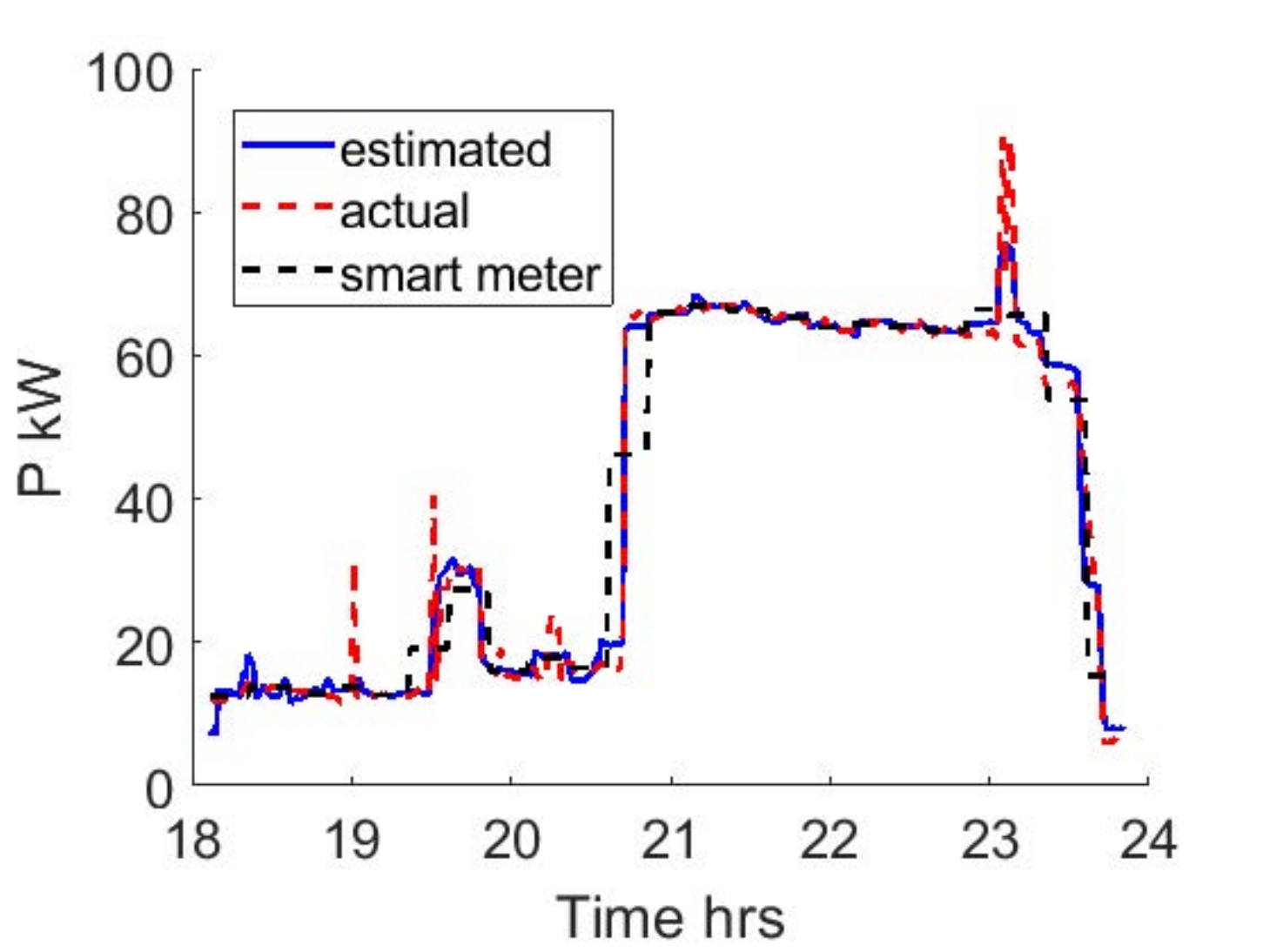}
		\centerline{(a)}
		\includegraphics[width=.8\linewidth]{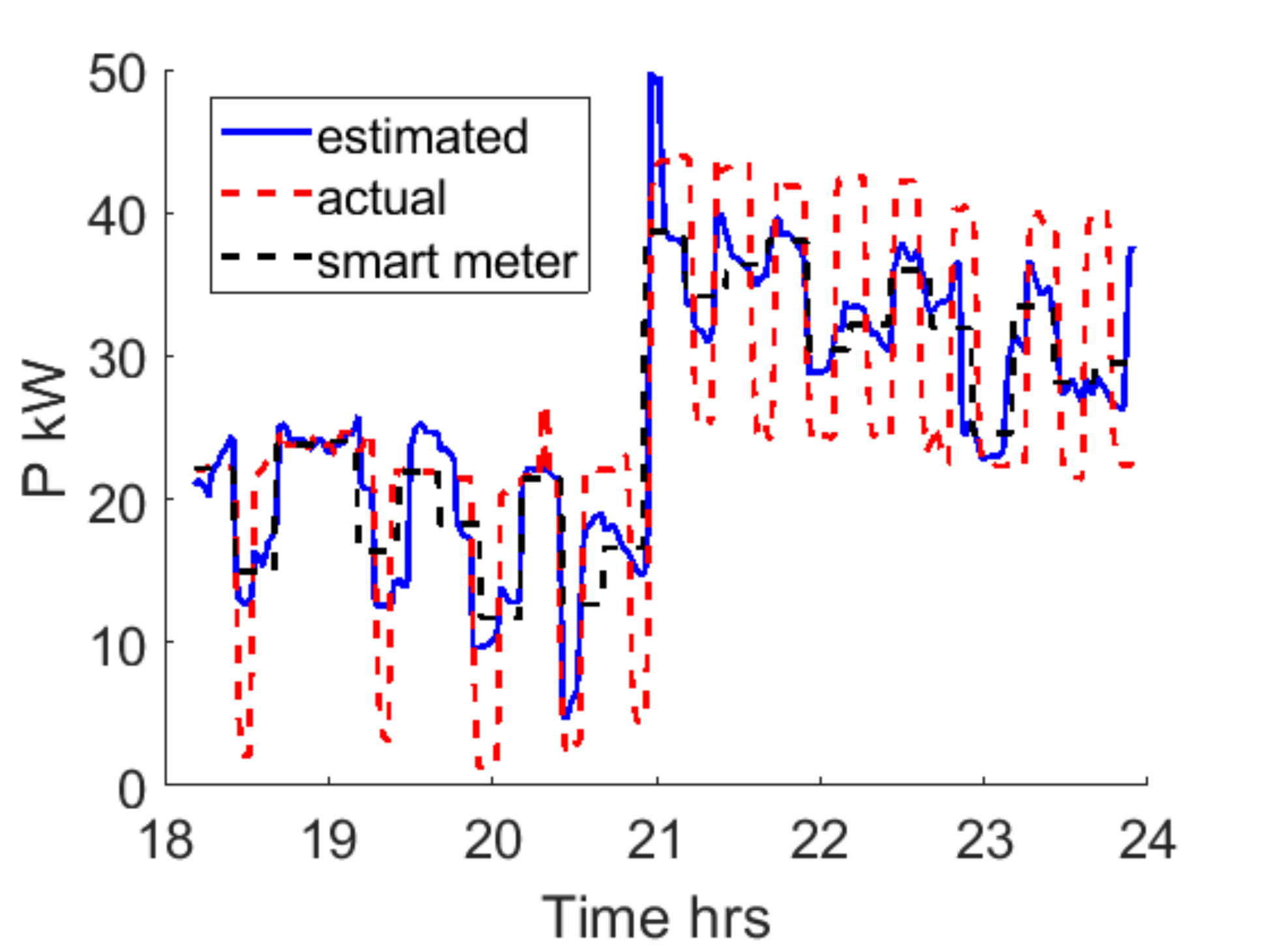}
		\centerline{(b)}
		\caption{{Recovered load data for one residential home with an EV charging event in Test Case 1 for the (a) winter and (b) summer night-time datasets.}}
		\label{fig:ev_30house}}
\end{figure}
We have tested the original recovery algorithm \eqref{eqn:multi_batch} using the 30-house system and the simplified one \eqref{eqn:multi_batch2} using the 100-house system. Their performances in detecting night-time EV charging events 
and recovering the day-time BTM solar output 
have been evaluated. Actual load data during both winter and summer seasons have been considered to demonstrate the impact of large periodic appliances as discussed in Remark \ref{rmk:hvac}.
To set up the optimization problem,  the sparsity coefficient $\lambda$ was fixed at 0.05 according to the $\ccalO(1/\sqrt{T})$ scaling. The smart meter error bounds $\boldsymbol{\mathcal{E}}$ were determined using the aforementioned metering accuracy as $0.2\%$ error for each entry. As for the D-PMU measurements, the error bounds $\bbE$ were set to be $0.2\%$ per measurement base on analyzing the linear approximation error of \eqref{s0_lin_eqn}. {Moreover, the smart meter measurements for the 30-house system are assumed to \textit{asynchronous}, while the 100-house system is assumed to be synchronous. Hence, we fix the measurement matrix $\bbH$ in \eqref{agg_eqn} for the 30-house system according to the average loading. Meanwhile, this matrix has been updated using the latest smart meter data in the 100-house system (cf. Remark \ref{rmk:agg})}. 
Recovery algorithms \eqref{eqn:multi_batch} and \eqref{eqn:multi_batch2} are implemented using the CVX toolbox \cite{cvx} with the MOSEK solver in the MATLAB\textsuperscript{\textregistered} R2020b simulator on a desktop with Intel\textsuperscript{\textregistered} Core\texttrademark~i5 CPU @ 3.40 GHz and 16 GB of RAM. 
%
\begin{figure}[t!]
	{
		\centering
		\includegraphics[width=.8\linewidth]{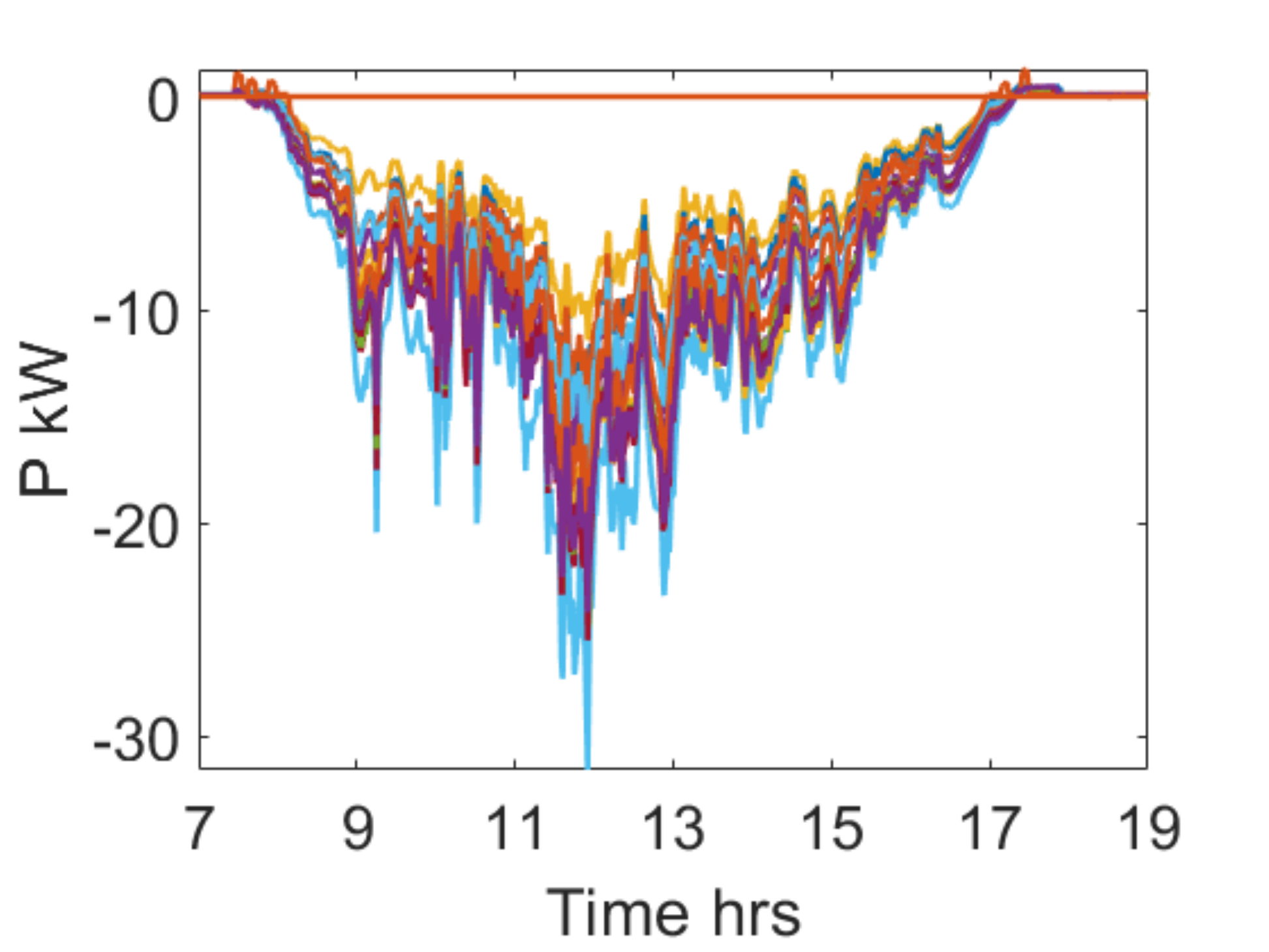}
		\centerline{(a)}
		\includegraphics[width=.8\linewidth]{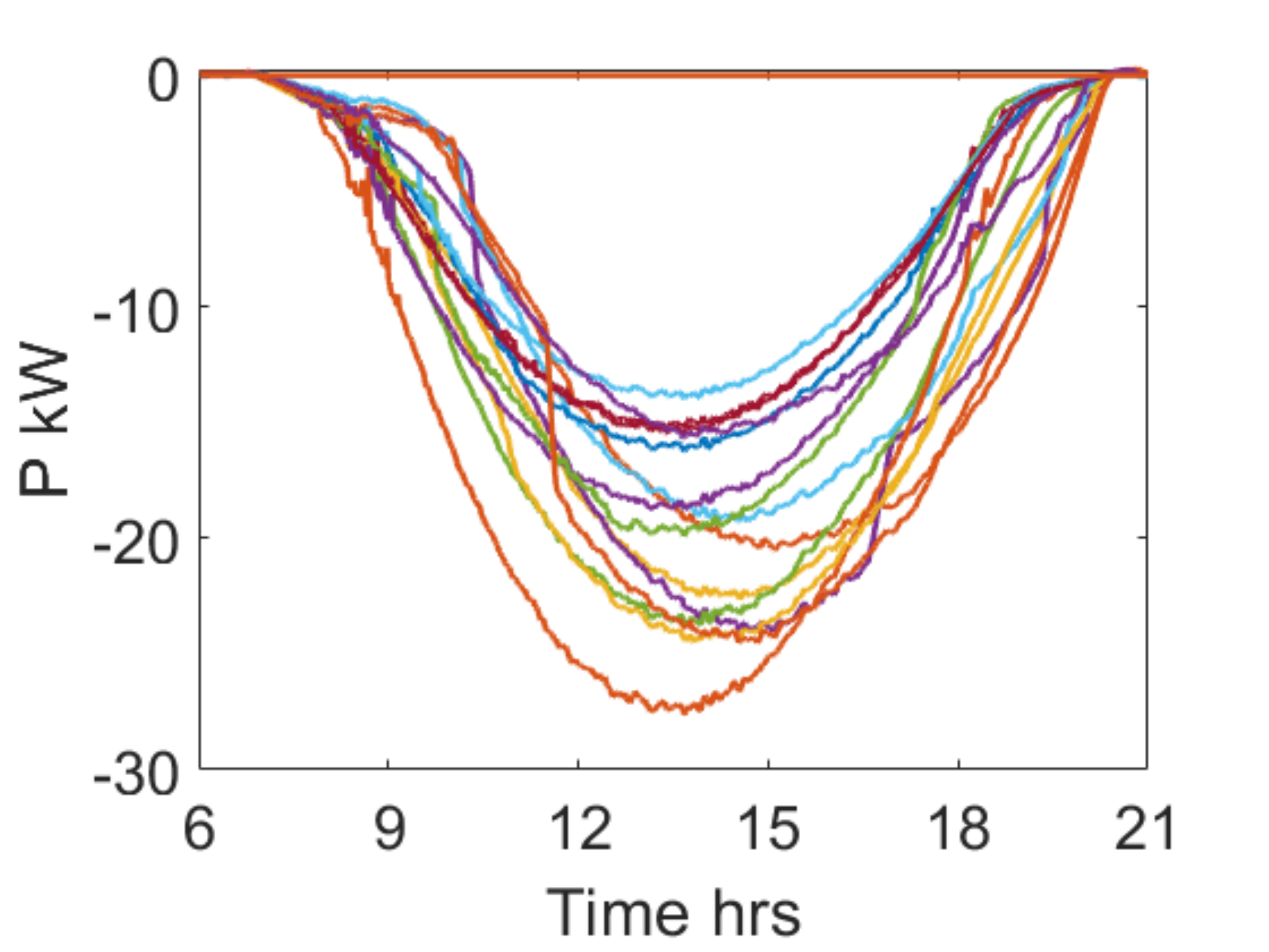}
		\centerline{(b)}
		\caption{{Ground-truth PV outputs from (a) the winter and (b) summer datasets in Test Case 2 for all 30 houses.}}
		\label{fig:ground_pv}}
\end{figure}


\textit{1) Test Case 1 on 30-house night-time data:} We first demonstrate the capability of recovery algorithm 
\eqref{eqn:multi_batch} in detecting EV events on the small feeder using the winter night-time dataset.  
We have increased the number of D-PMUs from only one at the feeder head to a total of $\kappa =4$ covering the full system. 
{Recall that the additional D-PMUs are placed at lateral heads.} The performance based on only smart-meter data is also included as denoted by $\kappa =0$ (no D-PMUs). The EV start/end events are detected by selecting a threshold value as a percentage of the EV power rating, and a successful detection is declared only if both the actual house and time instance have been correctly identified. To evaluate the detection performance, the receiving operating curve (ROC) is plotted in Fig.~\ref{fig:roc_30house}(a) by comparing the false positive rate (FPR) versus the true positive rate (TPR) for the varying detection threshold. Thus, the closer the ROC is to the top left corner (TPR = 1 and FPR=0), the better the detection performance is. To better visualize the ROCs, we  have smoothed the curves using an exponential function fitting of all the points. 
For the case of $\kappa=0$, Fig.~\ref{fig:roc_30house}(a) indicates that it is only possible to detect 
{$8.7\%$} of EV events with only smart meter data. This is exactly due to its low temporal resolution under the 15-minute averaging. By incorporating the total feeder power measurements using just one D-PMU, more than {$94\%$} 
of EV events can be correctly detected. 
Additional D-PMUs can further improve the TPR to detect almost $99\%$ of EV events, thanks to the their high temporal resolution. For all D-PMU cases ($\kappa \geq 1$) the FPR is very small, since regular household appliances have much lower power ratings than EVs and it is unlikely to mistakenly declare their activities for an EV event. {We have selected one residential house to illustrate the estimated profile {from 4 D-PMUs} during the night-time EV charging session, as shown in Fig.~\ref{fig:ev_30house}(a). Compared to the smart meter data, our estimated profile well matches the EV start/end charging time}. Hence, by using just a few D-PMUs, recovery algorithm \eqref{eqn:multi_batch} 
can effectively improve the spatial observability of the grid-edge DERs by identifying the exact location of EV typed activities in the feeder system. 
\begin{figure}[t!]
	{
		\centering
		\includegraphics[width=.8\linewidth,trim= 0 0 50 0, clip]{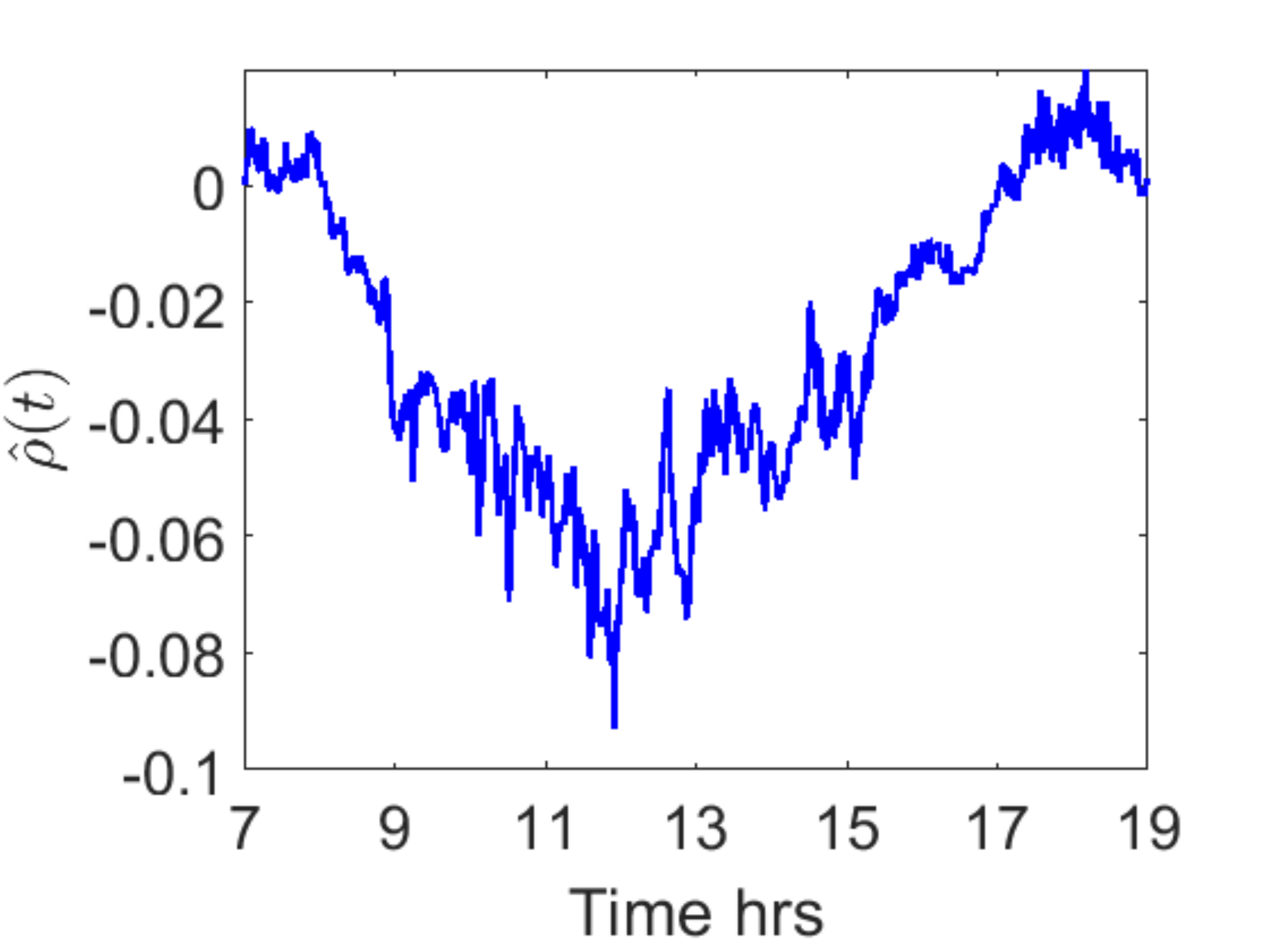}
		\centerline{(a)}
		\includegraphics[width=.8\linewidth, trim=50 20 100 0,clip]{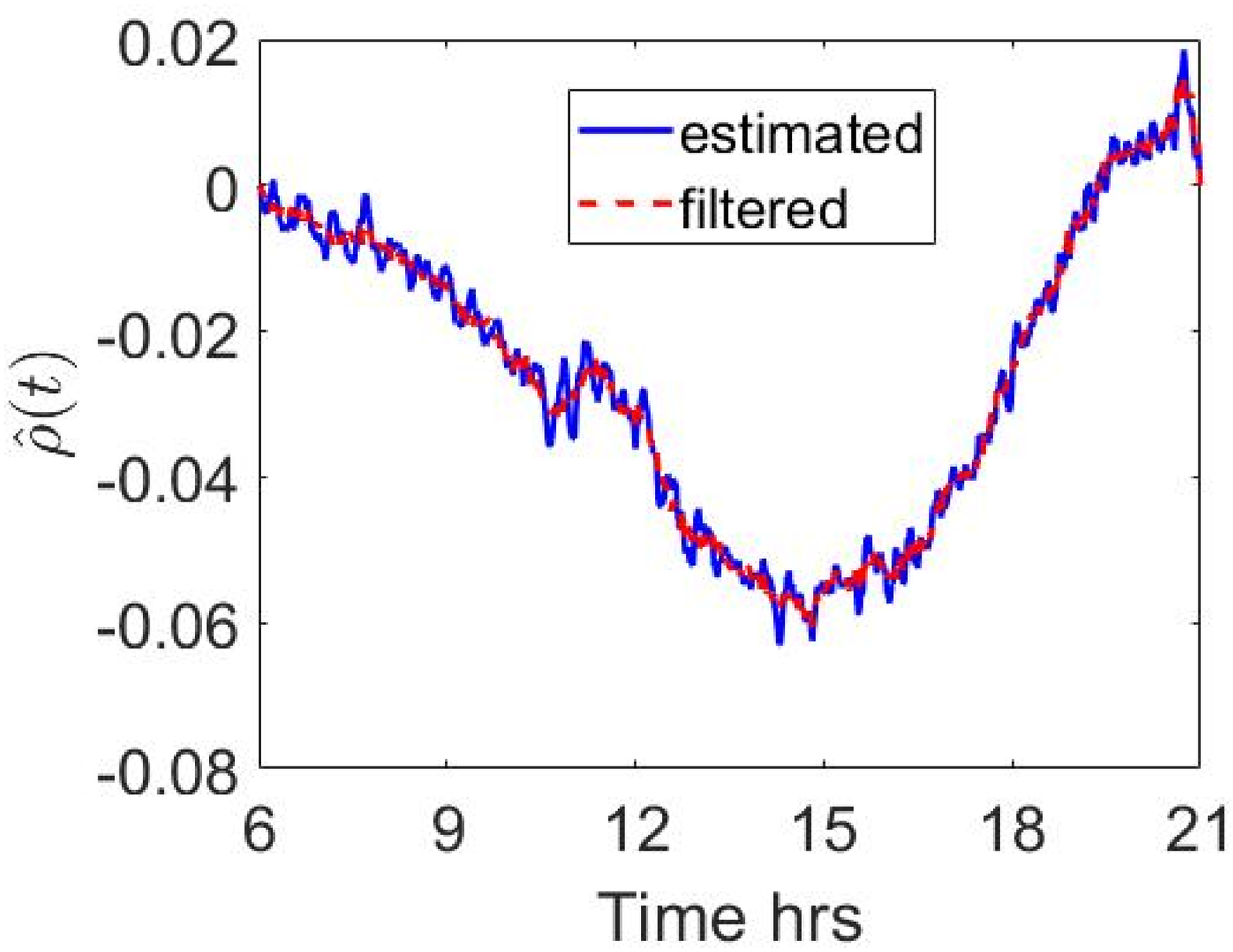}
		\centerline{(b)}
		\caption{{Recovered temporal patterns for the (a) winter and (b) summer day-time datasets in Test Case 2.}}
		\label{fig:rsv}}
\end{figure}

We further test the 
small feeder using a summer night-time dataset, with the ROCs plotted in  Fig.~\ref{fig:roc_30house}(b). As mentioned in Remark \ref{rmk:hvac}, HVAC loads in the summer violate the data modeling assumption for our proposed method, and thus have affected the detection performance. 
Compared to the winter results, the ROCs in Fig.~\ref{fig:roc_30house}(b) show reduced detection rates for all cases of $\kappa$. For one D-PMU, only 
{$74.6\%$} of EV events can be correctly identified, decreased from 
{$94\%$} earlier on.  Additional D-PMUs have again enhanced the accuracy to over 
$80\%$, but all cases experience an increase of FPR to be around $0.1$. This rising probability of false alarms is due to the high power rating of HVACs, which causes their activities to be declared as EV events by mistake. {One estimated load profile during an EV start event (around 20:55) is also illustrated in Fig.~\ref{fig:ev_30house}(b). Compared to the winter recovery, the summer one again confirms the capability of identifying EV events, yet periodic HVAC activities are much more difficult to recover.} Overall, our proposed method jointly utilizing D-PMU and smart meter data has attained enhanced spatial observability at the grid-edge. 
%
%
%
%
\begin{figure}[t!]
	{
		\centering
		\includegraphics[width=.8\linewidth,trim=50 10 50 10, clip]{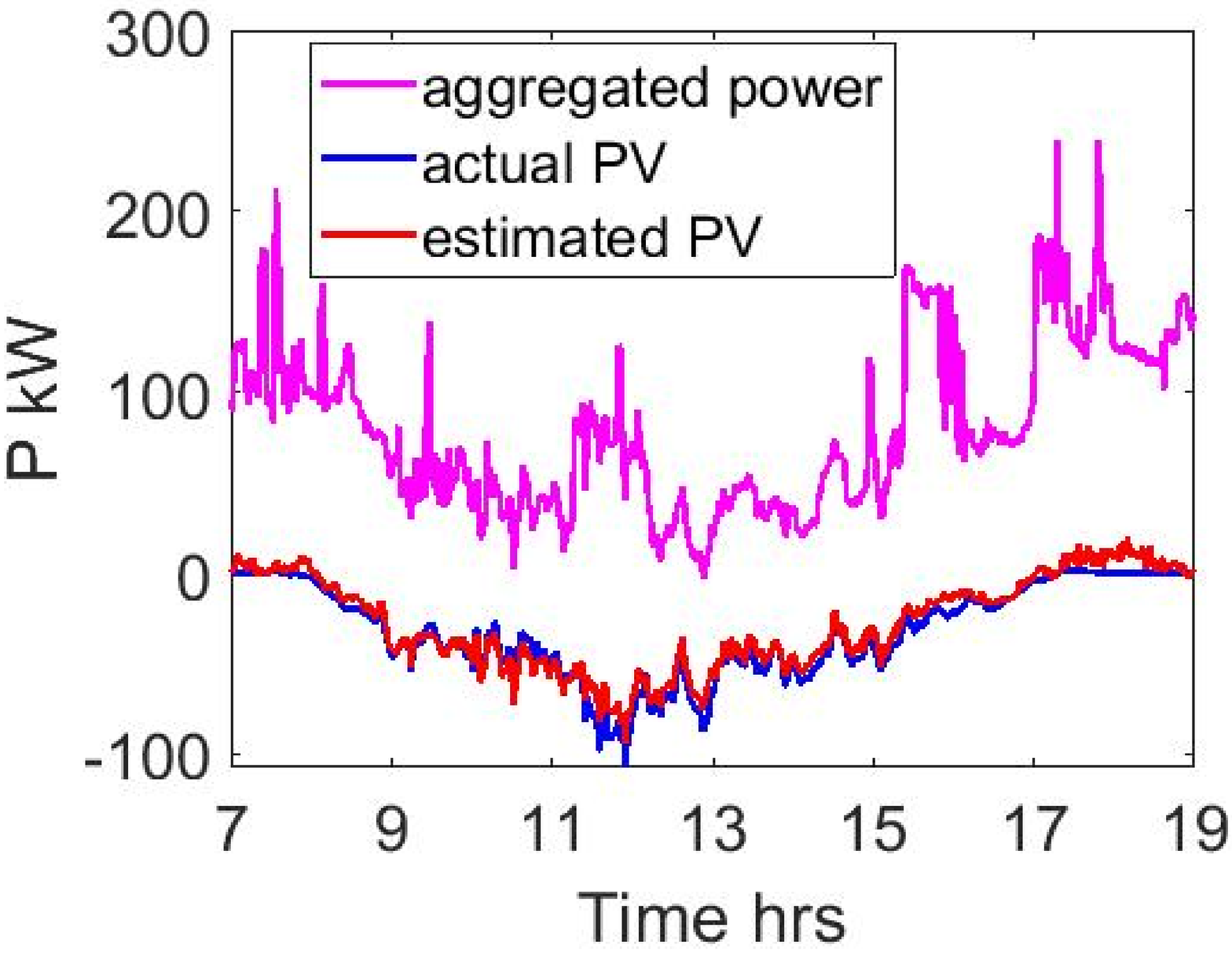}
		\centerline{(a)}
		\includegraphics[width=.8\linewidth]{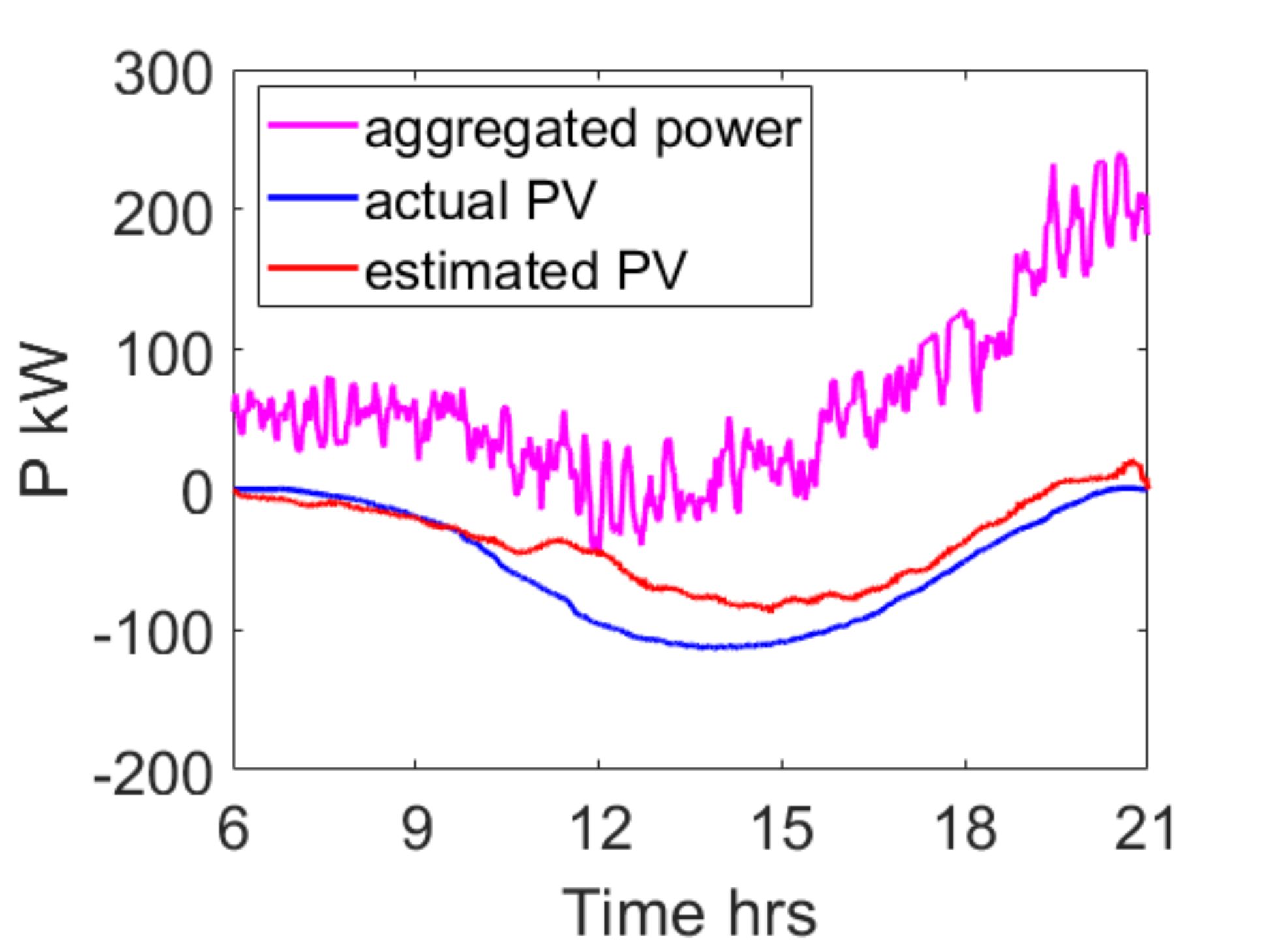}
		\centerline{(b)}
		\caption{{Estimated solar PV outputs from the aggregated D-PMU power data for the (a) winter and (b) summer datasets in Test Case 2.}}
		\label{fig:agg_pv}}
\end{figure}

\textit{2) Test Case 2 on 30-house day-time data:} 
We also test recovery algorithm \eqref{eqn:multi_batch} 
for recovering the BTM solar profile. The time window from hours 7:00-19:00 was used for the winter season and 
hours 6:00-21:00 for summer. These hours cover the entire day-time periods from sunrise to sunset. Fig.~\ref{fig:ground_pv}(a)-(b) plots the ground-truth PV outputs at all the houses 
for both datasets, respectively. The winter solar outputs exhibit high level of variability possibly due to cloud effects, while the summer ones tend to follow a smooth pattern of perfect solar irradiance. Due to the computer memory limitations, the recovery problem \eqref{eqn:multi_batch} has been solved for shorter time periods of 5-6 hours, with the (normalized) temporal patterns $\hat{\bbrho}$ recovered from the low-rank matrix solutions (as discussed in Remark \ref{rmk:BTM}) plotted in Fig.~\ref{fig:rsv}. 
The recovered winter solar pattern in Fig.~\ref{fig:rsv}(a) well matches the ground-truth profiles and even captures their fast variations around the noon hours for example. 
The summer solar pattern in Fig.~\ref{fig:rsv}(b) 
also well follows the parabolic trend of the ground-truth data, while showing an additional oscillation at a periodicity of {10-35 minutes}. 
As discussed in Remark \ref{rmk:hvac}, the periodic HVAC activities would also affect the low-rank component, giving rise to this periodic behavior in the summer temporal pattern. To deal with this issue, {we have used a simple band-pass filter to remove the frequency components in the periodicity range of 10-35 minutes from the estimated solar pattern (blue) to achieve a smoother pattern (red) as shown in Fig.~\ref{fig:rsv}(b). The recovered winter and filtered summer patterns} 
have been effective for estimating the total solar output from the D-PMU aggregated power data, 
as shown by Fig.~\ref{fig:agg_pv}. 
Using the processing discussed in Remark \ref{rmk:BTM}, we scaled the temporal pattern to estimate the total BTM solar output from each aggregated power profile. 
While the winter solar estimation can perfectly recover the actual PV output, the summer one has some mismatch in the peak magnitude. Thanks to the capability of recovering the temporal pattern, the proposed method is useful for enhancing the temporal observability regarding the BTM solar generation. An important future direction will be utilizing historical data and direct solar PV measurements to improve the scaling of BTM solar estimation. 
\begin{figure}[t!]
	{
		\centering
		\includegraphics[width=.8\linewidth]{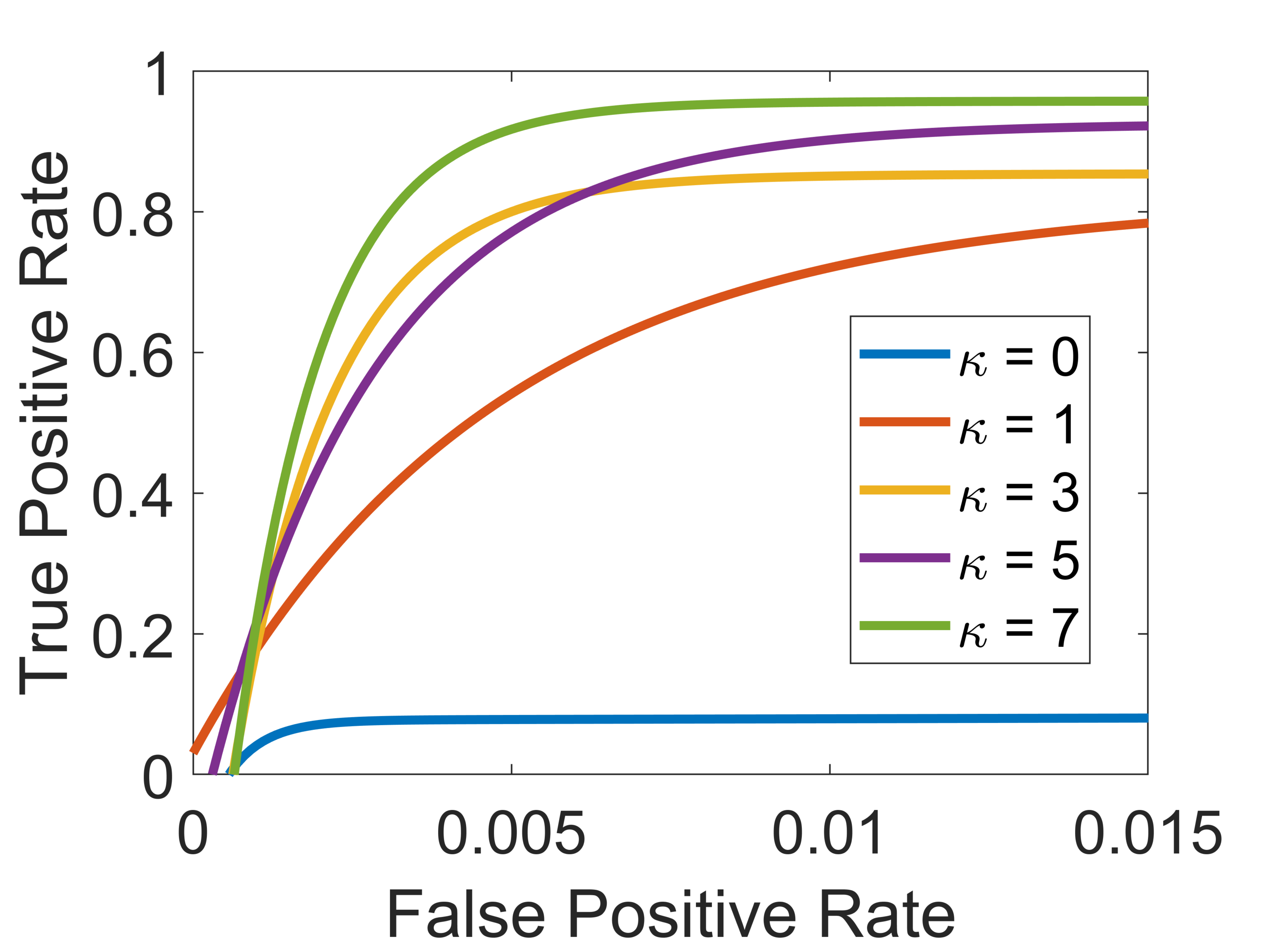}
		\centerline{(a)}
		\includegraphics[width=.8\linewidth]{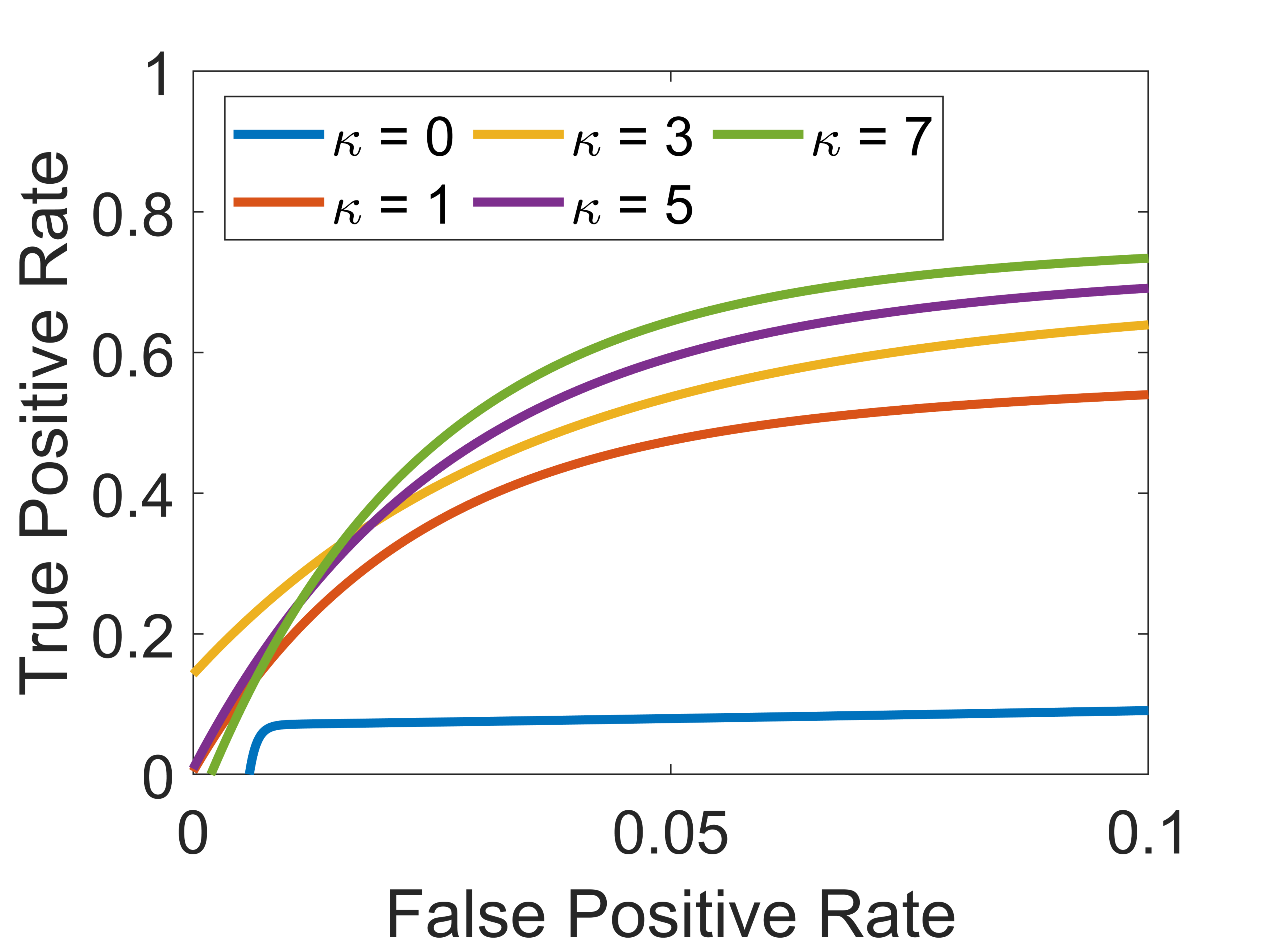}
		\centerline{(b)}
		\caption{ROCs of EV detection in Test Case 3 for the (a) winter and (b) summer night-time datasets with different number of D-PMUs.}
		\label{fig:roc_100house}}
\end{figure}

\textit{3) Test Case 3 on 100-house night-time data:} We demonstrate the computational improvement of the simplified method \eqref{eqn:multi_batch2} by applying it to the 111-node feeder in this test case. The settings are similar to Test Case 1, and both winter and summer datasets  are considered. The number of D-PMUs in the system has been varied from only one at the feeder head 
to a total of $\kappa = 7$. 
The ROCs comparing the FPR versus TPR are plotted in Fig.~\ref{fig:roc_100house}(a)-(b) 
for winter and summer 
respectively. 
Similarly to Test Case 1, the EV detection performance based on only smart meter data ($\kappa = 0$) is very poor due to its low temporal resolution. Less than $10\%$ of the EV events were correctly detected for both winter and summer datasets. By incorporating the D-PMU power data, Fig.~\ref{fig:roc_100house}(a) shows that the proposed method using \eqref{eqn:multi_batch2} can increase the detection rates to 
$78.3\% - 95.5\%$ for the winter dataset. Compared to the small system results in Fig.~\ref{fig:roc_30house}(a), a larger number of D-PMUs is needed here to maintain the high identification accuracy due to the increase of house locations. 
Similar observations can be made for the summer EV detection shown in Fig.~\ref{fig:roc_100house}(b), where the maximum TPR ranging from 
$54\% - 73.4\%$ is lowered than the values in Fig.~\ref{fig:roc_30house}(b).  Comparing between the two plots in Fig.~\ref{fig:roc_100house} again reveals the effects of HVAC loads in increasing the likelihood of false alarms (larger FPR values) during the summer recovery. Overall, the FPR is relatively small as most household appliances have smaller power ratings than EVs. Therefore, the large system tests demonstrate our simplified method \eqref{eqn:multi_batch2} can still effectively enhance the spatial observability regarding DER activities at the grid-edge.

\textit{4) Test Case 4 on 100-house day-time data:} We also used the simplified method \eqref{eqn:multi_batch2} for recovering the BTM solar profile. The time windows and key steps all follow from Test Case 2. 
The procedure in Remark \ref{rmk:BTM} was used to estimate the total BTM solar outputs as plotted in Fig.~\ref{fig:agg_pv_100house}. {A band-pass filter has again been implemented to smooth the recovered summer pattern in Fig.~\ref{fig:agg_pv_100house}(b).} 
Clearly, the recovered patterns well capture the trend of the actual solar PV outputs. Similar to Test Case 2, the recovered winter pattern can retain the fast solar transients while the summer one shows good match with the BTM PV output. Hence, the simplified method \eqref{eqn:multi_batch2} using a rank-one component maintains the capability of effectively recovering the underlying temporal pattern in the residential loads. As a result, it has attained good accuracy in estimating the total BTM solar output from the aggregated power measurements in distribution systems.

\begin{figure}[t!]
	{
		\centering
		\includegraphics[width=.8\linewidth]{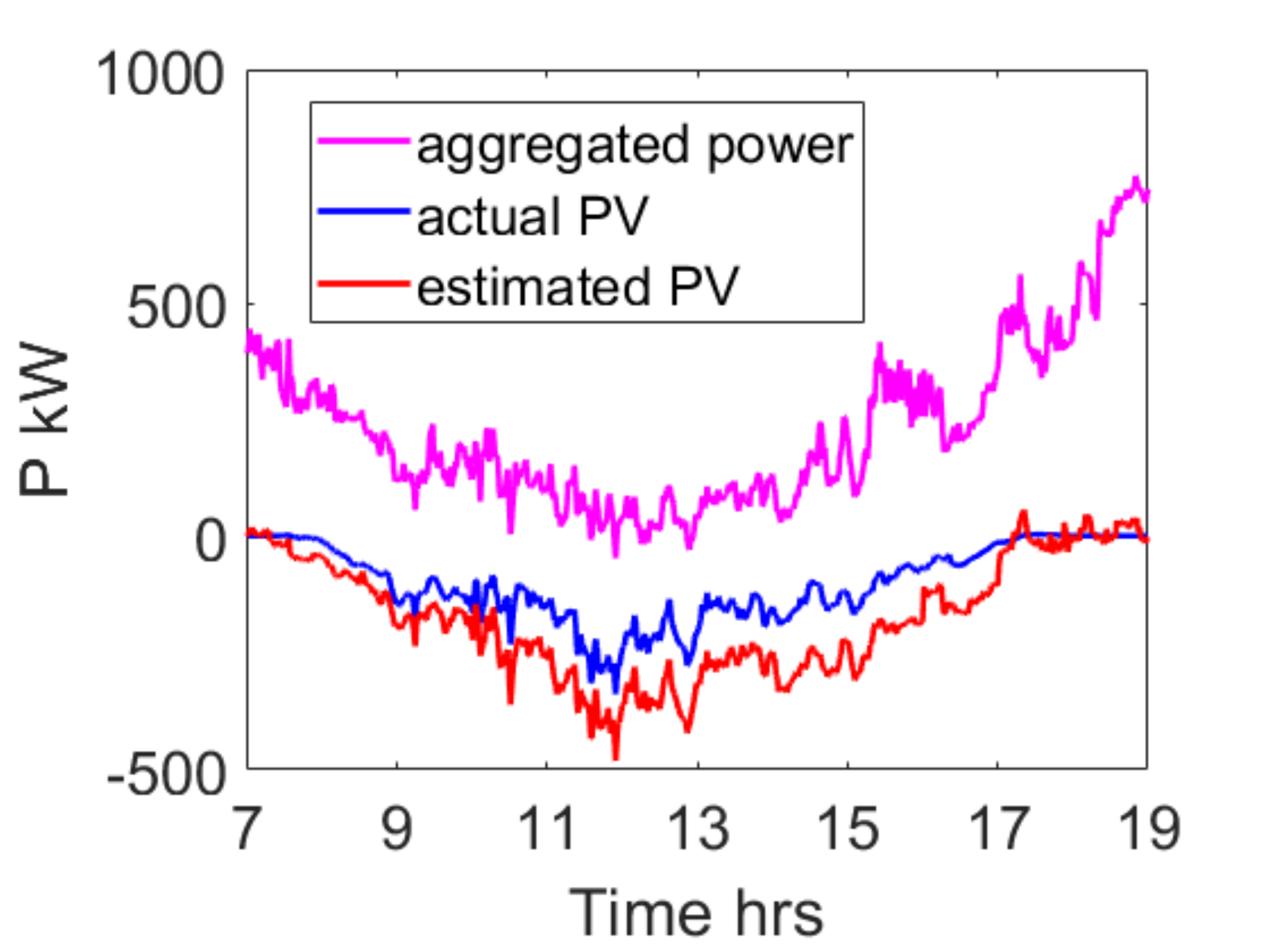}
		\centerline{(a)}
		\includegraphics[width=.8\linewidth]{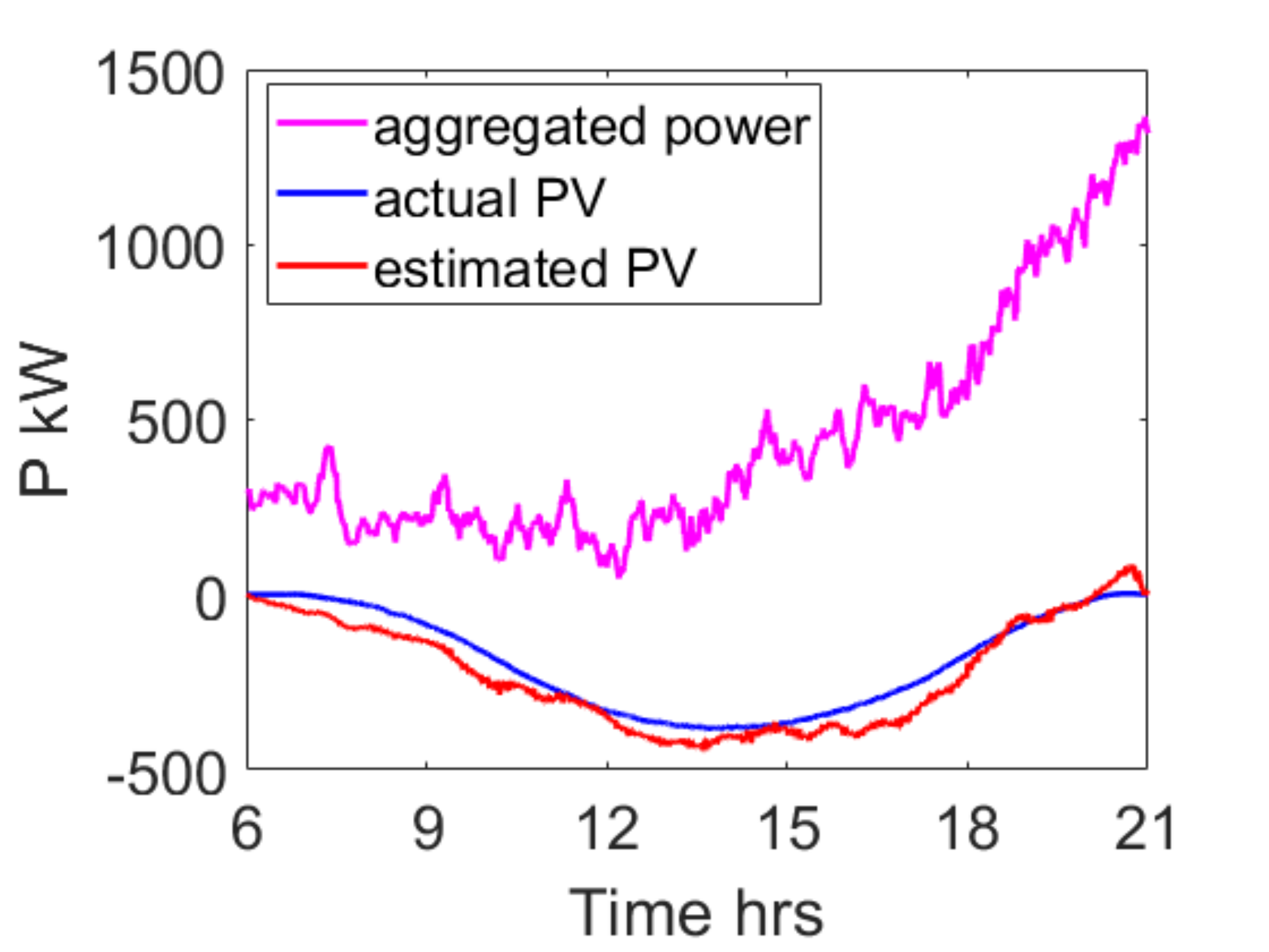}
		\centerline{(b)}
		\caption{{Estimated solar PV outputs from the aggregated D-PMU power data for (a) winter and (b) summer datasets in Test Case 4.}}
		\label{fig:agg_pv_100house}}
\end{figure}

\vspace*{3mm}
Finally, Table \ref{table:runtimes} lists the average runtime for the simplified formulation \eqref{eqn:multi_batch2} versus the number of D-PMUs $\kappa$ for a 4-hour time period and the 
larger feeder. The runtime grows gradually with  $\kappa$. Hence, additional D-PMUs can attain better monitoring performance (as shown by Test Case 3) at the cost of a slower runtime. {Additionally, the runtime is affected by the length of load data due to the increasing problem dimension. We plan to develop adaptive online implementation in future to address this computational issue.}

\begin{table}[t!]
\centering
\caption{Average runtimes of \eqref{eqn:multi_batch2} for varying number D-PMUs.}  \label{table:runtimes}
\begin{tabular}{c c}
\hline
$\kappa$ & Runtime (seconds) \\
\hline
1 & 45.53\\
3 & 111.77\\
5 & 195.52\\
7 & 421.22 \\
\hline
\end{tabular}
\end{table}


%
%


\section{Conclusions} \label{sec:con}

This paper developed a spatio-temporal learning approach that can jointly harness the respective strengths of D-PMU and smart meter measurements to enhance the observability of DERs. 
To tackle the lack of measurements, we exploited the underlying characteristics of residential loads and connected DERs. Activities of EVs and other appliances are considered to be infrequent events, and a group L1-norm is used to enforce \textit{jointly sparse} changes of active/reactive power. The strong correlation of solar generation within a feeder is modeled by a low-rank component, for which a nuclear norm is used. These two regularization terms have led to a convex recovery problem with linear constraints, which was further simplified by fixing the spatial scaling using the PV capacity information. Solutions to the recovery problems have been utilized to locate EV activities across the feeder and to infer the total BTM solar generation. Numerical tests using real-world load data on representative feeder systems have demonstrated the effectiveness of the proposed methods for monitoring DERs, although performance degradation has been observed in the presence of periodic HVAC loads. 

Interesting future research directions open up for this work. We are currently working on improving the solar pattern recovery by utilizing additional data sources such as direct solar metering. Furthermore, we plan to investigate efficient algorithms for accelerated computation time and to incorporate the identified load characteristics to the development of modeling and control approaches for distribution systems.

%
\bibliographystyle{IEEEtran}
\itemsep2pt
\bibliography{ref}

\newpage
\section*{Biographies}
\vspace{-10cm}
\begin{IEEEbiography} [{\includegraphics[width=1in,height=1.25in,keepaspectratio]{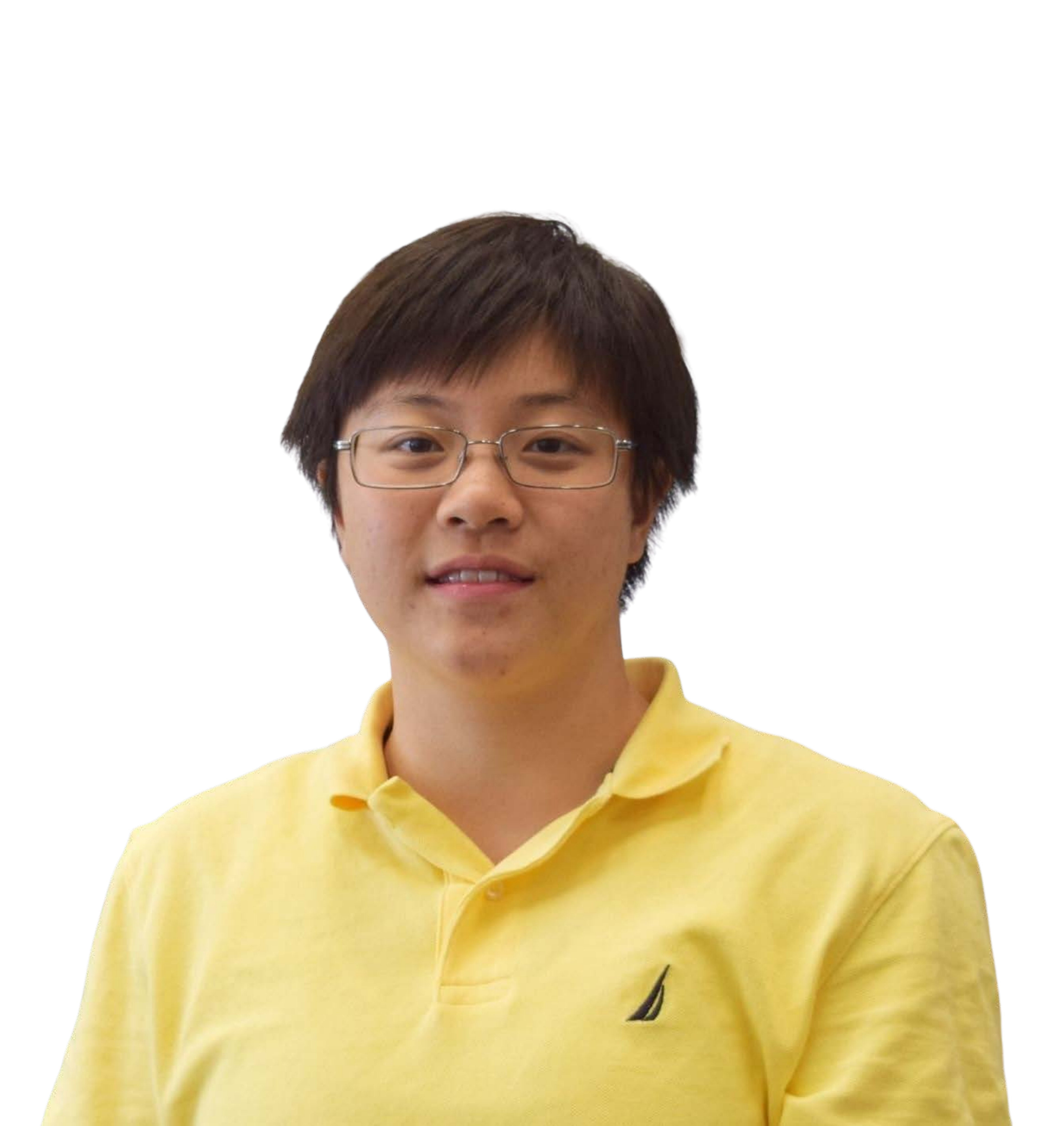}}]
{Shanny Lin} (S'20) received the B.S. degree in electrical engineering from Rensselaer Polytechnic Institute, Troy, NY, USA in 2018, and the M.S. degree from The University of Texas at Austin (UT Austin), Austin, TX, USA, in 2020. She is currently towards the Ph.D. degree at UT Austin. Her research interests include physics-aware and risk-aware machine learning for power distribution system operations. 
\end{IEEEbiography}
\vspace{-10cm}
\begin{IEEEbiography}[{\includegraphics[width=1in,
		height=1.25in,keepaspectratio]{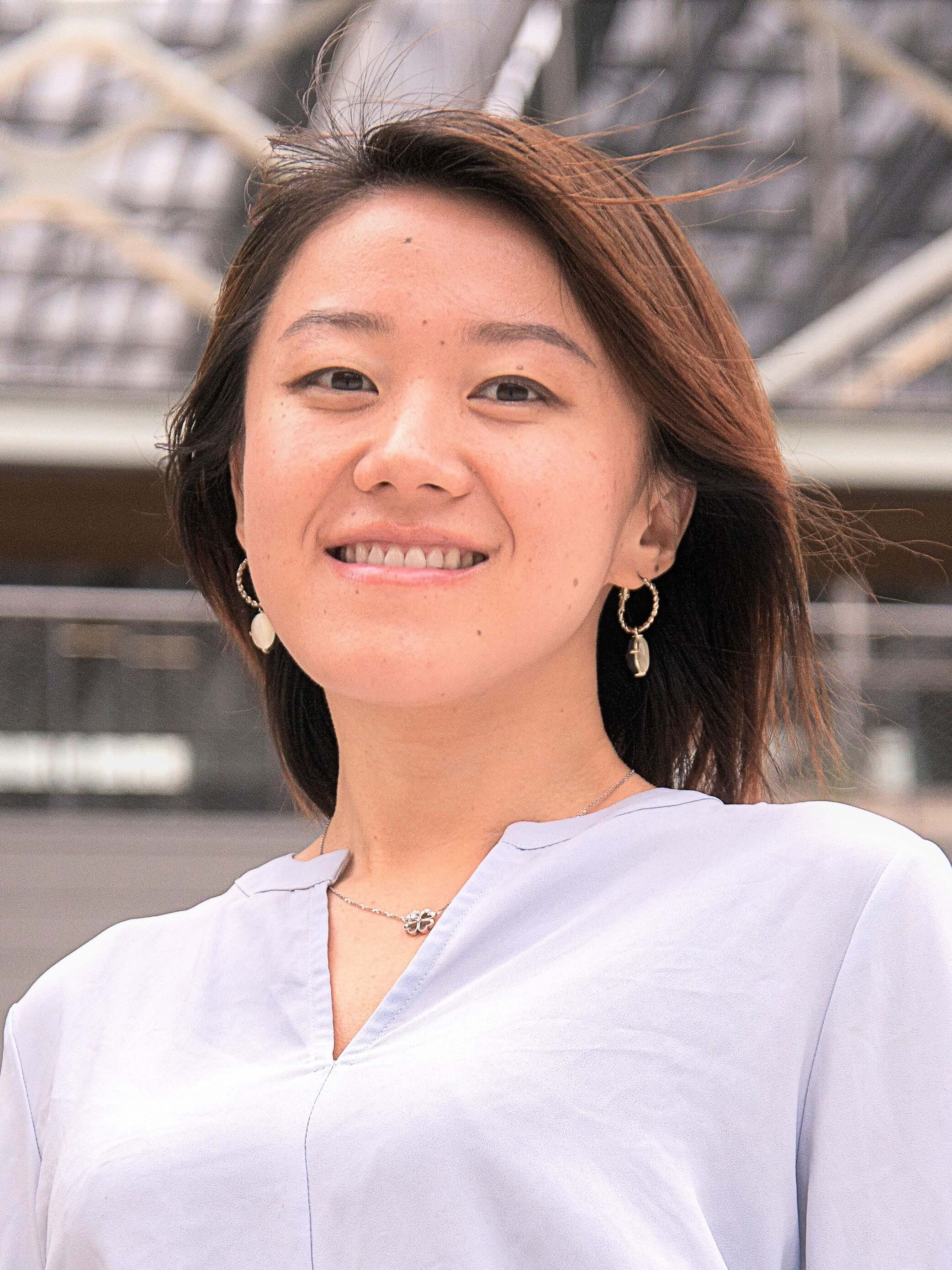}}]{Hao Zhu} (M'12--SM'19) is an Assistant Professor of Electrical and Computer Engineering (ECE) at The University of Texas at Austin. She received the B.S. degree from Tsinghua University in 2006, and the M.Sc. and Ph.D. degrees from the University of Minnesota in 2009 and 2012. From 2012 to 2017, she was a Postdoctoral Research Associate and then an Assistant Professor of ECE at the University of Illinois at Urbana-Champaign. Her research focus is on developing innovative algorithmic solutions for problems related to learning and optimization for future energy systems. Her current interest includes physics-aware and risk-aware machine learning for power system operations, and designing energy management system accounting for cyber-physical coupling. She is a recipient of the NSF CAREER Award and the Siebel Energy Institute Seed Grant Award, and also the faculty advisor for three Best Student Papers awarded at the North American Power Symposium. She is currently a member of the IEEE Power \& Energy Society (PES) Long Range Planning (LRP) Committee. 
\end{IEEEbiography}

\end{document}